\documentstyle[agupp,psfig]{article} 
\lefthead{HELMSTETTER and SORNETTE}
\righthead{Epidemic Models of Earthquake Aftershocks}	
\paperid{2001JB001580}
  \authoraddr{Agn\`es Helmstetter, Laboratoire de G{\'e}ophysique
Interne et Tectonophysique,
   Observatoire de Grenoble, Universit{\'e} Joseph Fourier, BP 53X, 38041
   Grenoble Cedex, France. (e-mail: ahelmste@obs.ujf-grenoble.fr)}
\authoraddr{Didier Sornette, Laboratoire de Physique de la
Mati\`{e}re Condens\'{e}e,
  CNRS UMR 6622 Universit\'{e} de Nice-Sophia Antipolis, Parc Valrose,
06108 Nice, France,
and Department of Earth and Space Sciences and Institute of
Geophysics and Planetary Physics, University of California, Los Angeles,
California. (e-mail: sornette@naxos.unice.fr)}

\input{update.tex}
\begin{document}
\title{Sub-critical and Super-critical Regimes in Epidemic Models of
Earthquake Aftershocks}
\author{Agn\`es Helmstetter}
\affil{Laboratoire de G{\'e}ophysique Interne et Tectonophysique,
  Observatoire de Grenoble, Universit\'e Joseph Fourier, France}
\author{Didier Sornette}
\affil{ Laboratoire de Physique de la Mati\`{e}re Condens\'{e}e, CNRS UMR 6622
Universit\'{e} de Nice-Sophia Antipolis, Parc Valrose, 06108 Nice, France and
Department of Earth and Space Sciences and Institute of
Geophysics and Planetary Physics, University of California, Los Angeles,
California 90095-1567}


\newcommand{\be}{\begin{equation}}
\newcommand{\ee}{\end{equation}}
\newcommand{\ba}{\begin{eqnarray}}
\newcommand{\ea}{\end{eqnarray}}
\newenvironment{technical}{\begin{quotation}\small}{\end{quotation}}


\begin{abstract}
We present an analytical solution and numerical tests of the
epidemic-type aftershock (ETAS) model for aftershocks, which describes
foreshocks, aftershocks and mainshocks on the same footing.
In this model, each earthquake of magnitude $m$ triggers aftershocks with
 a rate proportional to $10^{\alpha m}$.
The occurrence rate of aftershocks triggered by a single mainshock
 decreases with the time from the mainshock according to the modified
 Omori law $K/{(t+c)}^{p}$ with $p=1+\theta$. Contrary to the usual definition,
 the ETAS model does not impose an aftershock to have a magnitude smaller
 than the mainshock.
Starting with a mainshock at time $t=0$ that triggers aftershocks according
 to the local Omori law, that in turn trigger their own aftershocks and so on,
 we study the seismicity rate of the global aftershock sequence composed of
 all the secondary and subsequent aftershock sequences.
The effective branching parameter $n$, defined as the mean 
aftershock number triggered per event, controls the transition between 
a sub-critical regime $n<1$ to a super-critical regime $n>1$. 
A characteristic time $t^*$, function of all the ETAS parameters,
marks the transition from the early time behavior to the large time behavior.  
In the sub-critical regime, we recover and document the crossover from an  Omori exponent
$1-\theta$ for $t<t^*$ to
$1+\theta$ for $t>t^*$ found previously in [{\it Sornette and Sornette}, 1999]
for a special case of the ETAS model.
In the super-critical regime $n>1$ and $\theta>0$, we find a novel transition from an
Omori decay law with exponent  $1-\theta$ for $t<t^*$ to an explosive exponential increase of the
seismicity rate for $t>t^*$.
The case $\theta<0$ yields an infinite $n$-value.
In this case, we find another characteristic time $\tau$ controlling the crossover
 from an Omori law with exponent $1-|\theta|$ for $t<\tau$, similar to the local law, to an
 exponential increase at large times.
These results can rationalize many of the stylized facts reported for
aftershock and foreshock sequences, such as (i) the suggestion [{\it Liu}, 1984; {\it
Bowman}, 1997] that a small $p$-value may be a precursor of a large
earthquake, (ii) the relative seismic quiescence sometimes observed before
large aftershocks, (iii) the positive correlation between $b$ and $p$-values, 
(iv) the observation that great earthquakes are sometimes preceded by a decrease
 of $b$-value and (v) the acceleration of the seismicity preceding great earthquakes.

\end{abstract}
\begin{article}

\section{Introduction}
It is well known that the seismicity rate increases after a large earthquake,
for time period up to one hundred years [{\it Utsu et al.}, 1995],
and distances up
   to several hundred km [{\it Tajima and Kanamori}, 1985;
   {\it Steeples and Steeples}, 1996; {\it Kagan and Jackson},
   1998; {\it Meltzner and Wald}, 1999; {\it Dreger and Savage}, 1999].
   The  rate of the triggered events usually decays in time as the
modified Omori law
   $n(t) = K / {(t+c)}^p$, where the exponent $p$ is found to vary between
0.3 and 2 [{\it Davis and Frohlich}, 1991; {\it Kisslinger and Jones}, 1991;
{\it Guo and Ogata}, 1995; {\it Utsu et al.}, 1995] and is often close to 1
(see however [{\it Kisslinger}, 1993; {\it Gross and Kisslinger}, 1994]
for alternative decay laws such as the stretched exponential).


These triggered events are called aftershocks if their magnitude is smaller
than the first event. However, the definition of an aftershock contains
unavoidably a degree of arbitrariness because the qualification
of an earthquake as an aftershock requires the specification
of time and space windows. In this spirit,
several alternative algorithms for the definition of aftershocks have been
proposed [{\it Gardner and Knopoff}, 1974; {\it Molchan and
    Dmitrieva}, 1992] and there is no consensus.

 Aftershocks may result from several and not necessarily exclusive mechanisms
(see [{\it Harris}, 2001] and references therein): pore-pressure changes due
to pore-fluid flows coupled with stress variations, slow redistribution of
 stress by aseismic creep, rate-and-state dependent friction within faults,
  coupling between the viscoelastic lower crust and the brittle upper crust,
 stress-assisted micro-crack corrosion [{\it Yamashita and Knopoff}, 1987;
  {\it Lee and Sornette}, 2000],
  slow tectonic driving of a hierarchical geometry
with avalanche relaxation dynamics [{\it Huang et. al}, 1998], 
dynamical hierarchical models with heterogeneity, feedbacks and healing
[{\it Blanter et al.}, 1997], etc.

Since the underlying physical processes
are not fully understood, the qualifying time and space windows are more
based on common sense than on hard science.
Particularly, there is no agreement about the duration of the aftershock
sequence and the maximum distance between aftershock and mainshock.
If one event occurs with a magnitude larger than the first event,
it becomes the new mainshock and all preceding events are retrospectively
called foreshocks. Thus, there is no way to identify foreshocks
from usual aftershocks in real time.
There is also no way to distinguish aftershocks from individual earthquakes
[{\it Hough and Jones}, 1997].
The aftershock  magnitude distribution follows the Gutenberg-Richter
distribution with similar $b$-value as other earthquakes
[{\it Ranalli}, 1969; {\it Knopoff et al.}, 1982].
They have also similar rupture process.
Moreover, an event can be both an aftershock of a preceding large event,
and a mainshock of a following earthquake.
For example, the M=6.5 Big Bear event is usually considered as an aftershock
of the M=7.3 Landers event, and has clearly triggered its own aftershock
sequence. One can trace the difficulty of the problem
from the long-range nature of the interactions between
faults in space and time resulting in a complex self-organized crust.

In view of the difficulties in classifying sometimes an earthquake
as a foreshock, a mainshock or an aftershock, it is natural to investigate
models in which this distinction is removed and to study their possible
observable consequences. In this spirit,
the epidemic type  aftershock (ETAS) model introduced by {\it Kagan and Knopoff}
[1981, 1987] and {\it Ogata} [1988] provides
a tool for understanding the temporal clustering of the seismic
activity without distinguishing between aftershocks, foreshocks and mainshock events.
The ETAS model is a generalization of the modified Omori law, which takes
into account the secondary aftershocks sequences triggered by all events.
In this model, all earthquakes are simultaneously
mainshocks, aftershocks and possibly foreshocks. An observed ``aftershock''
sequence is in the ETAS model the result of the activity of all events
triggering events triggering themselves other events, and so on,
taken together.
The ETAS model aims at modeling complex aftershocks sequences and global
seismic activity. The seismicity rate is given by the superposition of aftershock
sequences of all events.
Each earthquake of magnitude $m$ triggers aftershock with a rate proportional
to $10^{\alpha m}$ with the same coefficient $\alpha$ for all earthquakes.
The occurrence rate of aftershocks decreases with the time from the mainshock
according to the modified Omori law $K /{(t+c)}^p$.
The background seismicity rate is modeled by a stationary Poisson
process with a constant occurrence rate $\mu$.
Contrary to the usual definition, the ETAS model does not impose an aftershock
to have a magnitude smaller than the mainshock.
This way, the same law describes both foreshocks, aftershocks and mainshocks.
This model has been used to give short-term probabilistic forecast of
seismic activity [{\it Kagan and Knopoff}, 1987; {\it Kagan and Jackson}, 2000;
 {\it Console and Murru}, 2001], and to describe the temporal and spatial clustering
of seismic activity [{\it Ogata}, 1988, 1989, 1992, 1999, 2001; {\it Kagan}, 1991;
{\it Felzer et al}, 2001].
Allthough the elementary results on the stability of the process
 have been known for many years [{\it Kagan}, 1991],
 no attempt has been made to study this model analytically in order to
 characterize its different regimes and obtain a deeper understanding
 of the combined interplay between the model parameters ($b$, $\alpha$, $p$, $K$, $c$
 and $\mu$) on the seismic activity. We stress below the contrast between previous works in the
 mathematical statistical literature and our results.

It should be noted that the ETAS model suffers from an important defect: it is
fundamentally a ``branching'' model [{\it Harris}, 1963; {\it Vere-Jones}, 1977],
with no ``loops''. What this means is that an event has a unique
``mother-mainshock'' and not several. In the real case, we can expect
that some events may be triggered by the combined loading and action at
distance in time and space of several previous earthquakes.
Hence, events should have several ``mothers'' in general.
This neglecting of ``loops'' is known in statistical physics as a
``mean-field'' approximation and allows us to simplify the analysis
while still keeping the essential physics in a qualitative way,
even if the details may not be precisely recovered quantitatively.

{\it Sornette and Sornette} [1999] studied analytically a particular case
of the ETAS model, in which the aftershock number does not depend on the
mainshock magnitude, {\it i.e.}, for $\alpha = 0$. Starting with one event
at time $t=0$ and considering that each earthquake generates an aftershock
sequence with a ``local'' Omori exponent $p=1+\theta$, where $\theta$ is
a positive constant, they studied  the decay law of the ``global'' aftershock
sequence, composed of all secondary  aftershock sequences.
They found that the global aftershock rate decays according to an Omori law
with an exponent $p=1-\theta$, smaller than the local one,
up to a characteristic time $t^*$, and then recovers the local Omori exponent
$p=1+\theta$ for time larger than  $t^*$.

Here, we generalize their analysis in the more general case
$\alpha >0$ of the ETAS model, which includes a realistic
magnitude distribution. We study the decay law of the global
aftershock sequence as a function of the model parameters (local
Omori law parameters and magnitude distribution). In addition
to giving more complete analytical results,
we present numerical simulations that test these predictions.
We also generalize the investigation and analysis into the
``super-critical'' regime. Indeed, depending on the branching
 ratio $n$, defined as the mean aftershock number
triggered per event, and on the sign of $\theta$,
three different regimes for the seismic rate $N(t)$ are found:
\begin{enumerate}
\item For $n<1$ (sub-critical regime),
we recover the results of [{\it Sornette and Sornette}, 1999],
{\it i.e.} we find a crossover from an  Omori exponent $p=1-\theta$
for $t<t^*$ to
$p=1+\theta$ for $t>t^*$.
\item For $n>1$ and $\theta>0$ (super-critical regime), we find a transition from an
Omori decay law with exponent  $p=1-\theta$ to an explosive exponential increase of
the seismicity rate.
\item In the case $\theta<0$, we find a transition from an Omori law with exponent
$1-|\theta|$ similar to the local law, to an exponential increase at large times,
with a crossover time $\tau$ different from the characteristic
time $t^*$ found in the case $\theta>0$.
\end{enumerate}

As we show below, these results can rationalize many properties of
aftershock and foreshock sequences.


\section{The model}

We assume that a given event (the ``mother'') of magnitude
$m_i \geq m_0$ occurring at time $t_i$
gives birth to other events (``daughters'') in the time interval
between $t$ and $t+dt$ at the rate
\be
\phi_{m_i}(t-t_i) = {K ~~10^{\alpha (m_i-m_0)}  \over
(t-t_i+c)^{1+\theta}} ~ H(t-t_i) ~ H(m_i-m_0 ) ,
\label{first}
\ee
where $H$ is the Heaviside function: $H(t-t_i) = 0$ for $t<t_i$ and
$1$ otherwise, $m_0$ is a lower bound magnitude below which no daughter
is triggered.

This temporal power law decay is Omori's law for the rate of aftershocks
following a main shock, albeit with the modification that we do not
specify that aftershocks (daughter earthquakes) have to be smaller
than the triggering event (mother earthquake). The exponential term
$10^{\alpha (m-m_0)}$ describes the fact that the larger the magnitude
 $m$ of the mother event, the larger is the number of daughters.
The exponent $p=1+\theta$ of the ``local'' Omori's law
has no reason a priori to be the same as the one measured
macroscopically which is usually found between $0.8$ and $1.2$ with
an often quoted median value $1$.
This is in fact the question we address\,: assuming the form (\ref{first})
 for the ``local'' Omori's law, is the global
Omori's law still a power law and, if yes, how does its exponent depend on $p$?
What are the possible regimes of aftershocks as a function of the parameters
of the model?

This model can be extended to describe the spatio-temporal distribution
 of seismic activity. Following {\it Kagan and Knopoff} [1981],
  we can introduce a spatial dependence in (\ref{first}) of the form
\be
\phi_{m_i}(t-t_i,\vec r-\vec r_i) = {K ~~10^{\alpha (m_i-m_0)}  \over
(t-t_i+c)^{1+\theta}} ~ \rho(\vec r-\vec{r_i})~ H(t-t_i) ~ H(m_i-m_0 ) ~,
\label{first2}
\ee
  where $\rho(\vec r-\vec{r_i})$ describes the probability distribution
for an earthquake occurring at position $\vec r_i$ to trigger an event
 an position $\vec r$.
 This term takes into account the spatial dependence of the stress induced
by an earthquake, and enable us to model the spatial distribution of aftershocks
 clustered close to the mainshock.
 In this paper, we restrict our analysis to the temporal ETAS model without
spatial dependence because we are mainly interested in describing the temporal
evolution of seismic activity. The complete model 
with both spatial and temporal dependence (\ref{first2})
has been studied in [{\it Helmstetter and Sornette}, 2002] to derive the joint
probability distribution of the times and locations of aftershocks including
the whole cascade of secondary aftershocks. When integrating the rate of
 aftershocks calculated for the spatio-temporal ETAS model over the whole
space, we recover the results given in this paper for the temporal ETAS model.
Therefore, the results given here for the temporal ETAS model
can be compared with real aftershock sequences when using all
aftershocks whatever their distance from the mainshock.

The model (\ref{first}) is a branching process because each daughter
has only one mother and not several, as shown in Figure
\ref{fig1branch}. As we said in the introduction, this
``mean-field'' assumption simplifies considerably the complexity of the
process and allows for an analytical solution that we shall derive in
the sequel.
The key parameter is the average number $n$ of daughter-earthquakes
created per mother-event.
Assuming that the distribution $P(m)$ of earthquake
sizes expressed in magnitudes $m$ follows the Gutenberg-Richter
distribution $P(m) = b~ \ln(10) ~ 10^{-b (m-m_0)}$,
the integral of $\phi_m(t)$ over time and over all magnitudes $m \geq m_0$
gives
\be
n \equiv \int_0^{+\infty} dt \int_{m_0}^{+\infty} dm~P(m)~ \phi_m(t)  =
n_0   ~ \int_0^\infty {dt \over (t+1)^{1+\theta}}~,
\label{second}
\ee
where
\be
n_0 \equiv {K \over  c^{\theta}} ~{b \over b - \alpha}~,
\label{demmfmlwwq}
\ee
which is finite for $b>\alpha$.
Three cases are analyzed below: $n<1, n=1$ and $n>1$. The case $n=1$ 
corresponds to an average conservation of the number of events and can be 
associated with a brittle elastic crust without dissipation.
The ``dissipative'' case $n<1$ can be interpreted as corresponding to a 
crust possessing a visco-elastic component and/or a partial coupling with a lower 
ductile layer, such that a part of the energy is released aseismically.
The case $n>1$ corresponds to a process in which an earthquake sequence
triggers an in-flow of energy from surrounding regions that may lead to a local
self-exciting amplification. It can also correspond to a coupling with other 
non-mechanical modes of energy storage, such as proposed in 
[{\it Sornette}, 2000b; {Viljoen et al.}, 2002] which can be triggered by an event and
feed the ensuing earthquake sequence for a while. Of course, the super-critical
process can only be transient and has to cross-over to another
regime.

The case $b<\alpha$ requires a special attention. In absence of truncation or
cut-off, it leads to a finite-time singularity due to the
interplay between long-memory and extreme fluctuations
[{\it Sornette and Helmstetter}, 2001]. However, it is more common to introduce
a truncation or
roll-off of the Gutenberg-Richter law at an upper magnitude.
We can for example use a Gamma distribution of energies, which is
 a power-law distribution tapered by an exponential tail.
In this case, the branching ratio has been calculated by {\it Kagan} [1991]
and is given by the approximate analytical expression 
valid for a corner magnitude $m_c$ significantly larger than $m_0$,
\be
n_0={K \over  c^{\theta}} ~{b \over b - \alpha} ~
{10^{b(m_c-m_0)}-10^{\alpha(m_c-m_0)} \over 10^{b(m_c-m_0)}-1} ~.
\label{demmlwwq}
\ee
 For a corner magnitude  $m_c \gg m_0$, and for $\alpha<b$, we recover
 the expression (\ref{demmfmlwwq}) for $n_0$ obtained for the Gutenberg-Richter
distribution without roll-off.

Note that $n$ is defined as the average  over all mainshock magnitudes
of the mean number of events triggered by a mainshock.
It is thus grossly misleading to think of the
branching ratio as giving the number of daughters to a given earthquake, because
this number is extremely sensitive to the specific value of its magnitude. Indeed,
the number of aftershocks to a given mainshock increases
exponentially with the mainshock magnitude as given by (\ref{first}),
 so that large earthquakes will have many more aftershocks than small earthquakes.
>From (\ref{first}) and (\ref{second}), we can calculate the mean
number of aftershocks $N(M)$ triggered directly by a mainshock of magnitude $M$
\be
N(M)=n {(b-\alpha) \over b} 10^{\alpha(M-m_0)}~.
\label{demtwwq}
\ee
As an example, take $\alpha=0.8$, $b=1$, $m_0=0$ and $n=1$. Then, a mainshock of magnitude
$M=7$ will have on average 80000 direct aftershocks, compared to only 2000 direct
aftershocks for an earthquake of magnitude $M=5$ and less than  0.2 aftershocks
for an earthquake of magnitude $M=0$.

When $\theta > 0$, $\int_0^\infty {dt \over (t+1)^{1+\theta}}=1/\theta$
and the branching ratio $n=n_0/\theta$ is finite.
In this regime, $n$ is an increasing function
of the rate $K$ and a decreasing function of $\theta$, $c$ and $b-\alpha$.

Even for $b>\alpha$ and $\theta >0$, the average number of daughters per mother can
be larger than one: $n > 1$. This regime corresponds to the super-critical
regime of branching processes [{\it Harris}, 1963; {\it Sornette}, 2000a] in
which the total number of events grows on average exponentially with time.
If $n <1$, there is less than one earthquake triggered per earthquake on average.
This is the sub-critical regime in which the number of events following
the first main shock decays eventually to zero. The critical
case $n=1$ is at the borderline between the two regimes.
In this case, there is exactly one earthquake on average triggered per
earthquake and the process is exactly at the critical point between death on
the long run and exponential proliferation.

There is another scenario, occurring for $\theta \leq 0$,
in which the seismicity blows up exponentially
with time. In this case,
the integral $\int_0^\infty {dt \over (t+1)^{1+\theta}}$
becomes unbounded. In principle, $n$ becomes infinite: this does
not invalidate the ETAS model
per se. It only reflects the fact that the calculation of an average number of
daughters per mother has become meaningless because of the anomalously slow
decay of the kernel $\phi(t)$. This mechanism is reminiscent of that leading
to anomalous diffusion and to aging in quenched random systems and spinglasses
(see [Sornette, 2000a] for an introduction). As in these systems, any
estimation of the averages depend on the time scale of study: due to 
the extremely
slow decay of $\phi(t)$, the number of daughters created beyond any 
time $t$ far
exceeds the number of daughters created up to time $t$.
Notwithstanding the decay,
its cumulative effect creates this dominance of the far future.
This regime is the opposite
of the situation where $\theta >0$ where most of the daughters are created at
relatively early times. Since the number of daughters born up to time $t$
is an unbounded increasing function of $t$, it is intuitively appealing,
as we show in the appendix,
that this regime should be similar to the super-critical regime $n>1$
discussed above in the case $\theta >0$.

Until now, we have discussed three issues related to the convergence of the ETAS 
sequences:
(i) the condition $\theta >0$ ensures convergence at large times;
(ii) the convergence at short times is obtained by the introduction of
the regularization constant $c$ in the generalized Omori's law;
(iii) the condition $\alpha < b$ is a necessary condition for the finiteness
of the number of daughters. Finally, we should stress the role of the
``ultra-violet'' cut-off $m_0$ on the magnitudes. In the ETAS model,
only earthquakes of magnitude $m \geq m_0$ are allowed to give birth
to aftershocks, while events of smaller magnitudes are lost for the epidemic
dynamics. If such a cut-off is not introduced 
and no cut-off is put on the Gutenberg-Richter toward small magnitudes, the
dynamics becomes completely dominated by the swarms of very tiny earthquakes, 
which individually has very low probability to generate aftershocks but become
so numerous that their collective effect becomes overwhelming in the dynamics. We would
thus have the unphysical situation in which a magnitude $7$ or $8$ earthquake
may be triggered by tiny earthquakes of magnitudes $-2$ or less. 
We stress that the introduction of such a cut-off $m_0$ is a simple way to 
prevent such a situation to occur, but it does not mean that small earthquakes
of magnitude below $m_0$ do not have their own aftershocks. It only means that 
such small earthquakes create aftershocks that can not participate in the 
epidemic process leading to significantly larger earthquakes; these small earthquakes
live their separate life. This is why they are not 
registered by the ETAS model. This formulation is of course only an end-member
of many possible regularization procedures, which are well-known to be an
ubiquitous requisite in mechanical models of rupture. An improvement of the ETAS model
would be for instance to replace this abrupt cut-off $m_0$ by introducing 
a roll-off in the Gutenberg-Richter law for the aftershocks with a characteristic corner magnitude
decreasing with the magnitude of the mother earthquake. This and other schemes will not
be explored here, as we want to analyze the simplest version possible.

We now describe briefly the connection with previous works in the mathematical statistics
literature.
As we said above, the model (\ref{first}) belongs to the general class of branching models
[{\it Moyal}, 1962; {\it Harris}, 1963].
The elementary results on the stability of the process, such as the condition $n < 1$,
have been known for many years, and go back to the origin
of the ETAS model as a special case (for discrete magnitudes) or extension (for
continuous magnitudes) of the class of ``mutually exciting point processes''
introduced in [{\it Hawkes}, 1971; 1972; {\it Hawkes and Adamapoulos}, 1973]. 
A  convenient mathematical overview is in Chapter 5 of Vere-Jones and Daley [1988], especially
Example 5.5(a) and associated exercises 5.5.2-5.5.6. 
For the
ETAS model, the equations governing the probability generating functional, the probability
of extinction within a given number of generations, the expectation measure
for the total population, the second factorial moment (related to the covariance
of the population) and their Fourier transform 
can be derived as special cases of results summarized there. In
particular, the process initiated with a single event at the origin corresponds
to the total progeny process for a general branching process model with
time-magnitude state space and a single ancestor at time $t=0$; Exercise 5.5.6
gives the equations of the above
cited variables for the case of fixed magnitudes (i.e., $\alpha =0$). This direct probabilistic
analysis in terms of generating functions
effectively  replaces the  Wiener-Hopf theory in the present paper and mentioned also 
in [{\it Hawkes}, 1971; 1972; {\it Hawkes and Oakes}, 1974]. However, there is not
explicit solutions given to these equations and there is
no discussion of the change of regime from an effective Omori's law
$1/t^{1-\theta}$ at early times to $1/t^{1+\theta}$ at long times, nor mention
of the interesting super-critical case, as done in the present work.

Hawkes [1971; 1972] and Hawkes and Adamapoulos [1973]
use what is in effect an ETAS model with an exponential
``bare'' Omori's law rather than the power law $1/(t+c)^{1+\theta}$ defined in 
(\ref{first}). Hawkes and Adamapoulos [1973] use it in 
an early study of earthquake data. 
The introduction of magnitudes is similar to the introduction of 
a marked process associated with a single point process [{\it Hawkes}, 1972]; 
however, the impact of magnitudes on the seismicity rate is 
assumed to be linear in [{\it Hawkes}, 1972] while it is multiplicative in the 
ETAS model. Our derivation presented in the
appendix of the solution of the ETAS model for the mean rate of 
earthquakes in terms of its Laplace transform recovers previous results.
For instance, equation (17) in [{\it Hawkes and Oakes}, 1974] is the same as our
equation (\ref{bghnglal}) in our Appendix (up to a factor $\beta$ stemming
from taking the cumulative number in [{\it Hawkes and Oakes}, 1974]).
The key factor $Q(\beta )$ in (\ref{jgmjslqlqllq})
corresponds to the quantity $G_1(0)$ in equation (5) of [{\it Hawkes}, 1972]. 
The link between Hawkes' ``mutually exciting point processes''
and branching processes was made explicit in [{\it Hawkes and
Oakes}, 1974].

Some average properties of the  ETAS model have been derived in
the Master thesis of P.A. Ramselaar (1990). Specifically, 
using the theory of Markov processes applied to branching processes,
Ramselaar (1990) proves that, in the supercritical regime $n>1$ 
(where $n$ is the average branching ratio
defined in (\ref{demmlwwq})), the average number of aftershocks stemming from
a common ancestor grows exponentially as $\sim e^{t/t^*}$ where $t^*$ is the
solution of $n R(c/t^*) =1$ and the function $R$ is defined in (\ref{hghigwnv}).
The solution of this equation $n R(c/t^*) =1$ for $t^*$ is the same as
our $t^*$ given by (\ref{ngnvl}) and the exponential growth of Ramselaar is therefore
the same as our result (\ref{gnalksd1}). We add on this asymptotic result, which is 
valid only at large times, by exhibiting
the solution for the aftershock decay at early times. In addition, contrary to the 
incorrect claim of Ramselaar (1990) that ``the Ogata earthquake process is critical
or supercritical but is never subcritical,'' we demonstrate that the subcritical 
regime exhibits a rich phenomenology.

\section{Analytical solution}

We analyze the case where there is an
origin of time $t=0$ at which we start recording the rate of
earthquakes, assuming that
the largest earthquake of all has just occurred at $t=0$ and somehow
reset the clock. In the
following calculation, we will forget about the effect of events
preceding the one at $t=0$
and count aftershocks that are created only by this main shock.

Let us call $N_m(t)$ the rate of seismicity at time $t$ and at 
magnitude $m$, that is, $N_m(t) dt dm$ is
the number of events 
in the time/magnitude interval $dt\, \times \,dm$. We define its
expectation $\lambda_m(t) dt dm \equiv {\rm E}[N_m(t)\,dt\,dm]$, as
the mean number of earthquakes occurring between
$t$ and $t+dt$ of magnitude
between $m$ and $m+dm$. $\lambda_m(t)$ is the
solution of a self-consistency equation that formalizes
mathematically the following
process\,: an earthquake may trigger aftershocks; these aftershocks may trigger
their own aftershocks, and so on. The rate of seismicity at a given
time $t$ is the
result of this cascade process. The self-consistency equation that sums up this
cascade reads
\ba
 \lambda_m(t) \equiv  {\rm E}[N_m(t)]
 & =& E\left[ \int_{m_0}^{\infty} dm'\int_{-\infty}^t d\tau~
\phi_{m'}(t-\tau) P(m) N_{m'}(\tau) \right] \\
  &=& \int_{m_0}^{\infty} dm'\int_{-\infty}^t d\tau~ \phi_{m'}(t-\tau) {\rm E}[N_{m'}(\tau)] \\
\,\,&=  & \int_{m_0}^{\infty} dm'\int_{-\infty}^t d\tau~  \phi_{m'}(t-\tau) \lambda_{m'}(\tau)\,. 
\label{third}
\ea
If there is an external source $S(t,m)$, it should be added to the right-hand-side of 
(\ref{third}).

The mean instantaneous rate $\lambda_m(t)$ at time $t$ is the sum over
 all induced rates
from all earthquakes of all possible magnitudes
that occurred at all previous times.  The rate of events at time $t$
induced per earthquake that occurred at an earlier time $\tau$ with magnitude
$m'$ is equal to $\phi_{m'}(t-\tau)$.
The term $P(m)$ is the probability that an event triggered by an earthquake
 of magnitude $m'$ is of magnitude $m$. We assume that this probability is
  independent of the magnitude of the mother-earthquake and is nothing
but the Gutenberg-Richter law.
This hypothesis can be easily relaxed if needed and $P(m)$ can be generalized into
$P(m|m')$ giving the probability that a daughter-earthquake is of magnitude $m$
conditioned on the value $m'$ of the magnitude of the mother-earthquake.
However, we do not pursue here this possibility as this hypothesis seems well-founded
empirically [{\it Ranalli}, 1969; {\it Knopoff et al.}, 1982].
The term $S(t,m)$ is an external source which is determined by the
physical process. We consider the case where a great earthquake occurs at the origin
  of time $t=0$ with magnitude $M$. In this case, the external source term is
\be
S(t,m)=\delta(t)~\delta(m-M)~,
\label{S(t,m)}
\ee
where $\delta$ is the Dirac distribution.
Other arbitrary source functions can be chosen.

 By construction of the kernel (\ref{first}), the integrals over the magnitudes
and over time factorize, which implies that the solution for $\lambda_m(t)$
can be searched as
\be
\lambda_m(t) = P(m) \lambda(t)~,  \label{ggnnlalaq}
\ee
which makes explicit the separation of the variables magnitude and time.
We stress that (\ref{ggnnlalaq}) is not
an assumption: it is the structure of the solution based on
the assumption made at the level of
equations (\ref{first}) and (\ref{third}). Therefore, the ETAS model
assumes that the time response and the magnitude response are independent.
In reality and more generality, we can envision that the rate of
activation of new earthquakes will depend on 1) the magnitude of the ``mother''
(which the ETAS model takes into account
multiplicatively in (\ref{first})), 2) on the magnitude of the
daughter (which is neglected in the ETAS model)
and 3) on the time since the mother was born. Rather than having a very
general kernel combining these three parameters nonlinearly,
equations (\ref{first}) and (\ref{third}) are based on an hypothesis of
independence between these different factors
that allows us to factorize them, leading to (\ref{ggnnlalaq}).

The problem is then to determine the functional form of $\lambda(t)$, assuming that
$\phi$ is given by (\ref{first}). The integral equation (\ref{third}) is a
Wiener-Hopf integral equation [{\it Feller}, 1971].
It is well-known [{\it Feller}, 1971; {\it Morse and Feshbach}, 1953]
that, if $\phi(\tau)$ decays no slower than an exponential, then
$\lambda(t)$ has an exponential tail $\lambda(t) \sim \exp[-r t]$ for large $t$ with
$r$ solution of $\int \phi(x) ~\exp[r x]~dx = 1$. This result implies that 
a global Omori's law cannot be obtained by the epidemic ETAS branching model
with, for instance, local exponential relaxation rates.
In the present case, $\phi(\tau)$ decays much slower than an exponential and a
different analysis is called for that we now present.
The solution of (\ref{third}) is derived in the Appendix and is summarized
in the following sections. For the sequel, it is useful to define
the characteristic time
\be
t^* \equiv c \left({n ~ \Gamma(1-\theta) \over |1-n|}\right)^{1 \over \theta}~,
\label{ngnvl}
\ee
where $\Gamma(x)$ is the Gamma function: $\Gamma(z) =\int_0^{\infty} 
du ~u^{z-1} ~e^{-u}$
which is nothing but $(z-1)!$ for positive integers $z$.

\subsection{The sub-critical regime $n<1$ and $\theta>0$}

An approximation is made in the analytical solution so that the 
results presented
below are only valid for $t\gg c$.

We define the parameter $S_0$ that describes the external source term
\be
S_0={(b-\alpha)\over b} 10^{\alpha(M-m_0)} ~.
\label{S02}
\ee

Two cases must be distinguished.

$\bullet$ For $c \ll t \ll t^*$, we get
\be
\lambda_{t<t^*}(t) \sim { S_0 \over \Gamma(\theta )(1-n) }~~~
{{t^*}^{-\theta} \over t^{1-\theta}} ~~~~~ \mbox{  for $c\ll t \ll t^*$  .}
\label{eadfgz1}
\ee

$\bullet$ For $t \gg t^*$, we obtain
\be
\lambda_{t>t^*}(t) \sim { S_0  \over \Gamma(\theta) (1-n)}~~
{ {t^*}^{\theta}\over t^{1+\theta}} ~~~~~ \mbox{  for $t\gg t^*$  ~.}
\label{eaz1}
\ee

We verify the self-consistency of the two solutions $\lambda_{t>t^*}(t)$
and $\lambda_{t<t^*}(t)$
by checking that $\lambda_{t>t^*}(t^*) = \lambda_{t<t^*}(t^*)$. In other words,
$t^*$ is indeed
the transition time at which the ``short-time''
regime $\lambda_{t<t^*}(t)$ crosses over to the ``long-time'' regime $\lambda_{t>t^*}(t)$.

The full expression of $\lambda(t)$ valid at all times $t \gg c$ is given by
\be
\lambda(t) = {S_0 \over 1-n}~{{t^*}^{-\theta} \over t^{1-\theta}}~
\sum_{k=0}^\infty (-1)^k {(t/t^*)^{k\theta} \over
\Gamma((k+1)\theta)}
\label{bhgknaalkal2}
\ee
Expression (\ref{bhgknaalkal2}) provides the solution that describes
the cross-over from the $1/t^{1-\theta}$ Omori's law (\ref{eadfgz1})
at early times to
the $1/t^{1+\theta}$ Omori's law (\ref{eaz1}) at large times.
The series $\sum_{k=0}^\infty (-1)^k {(t/t^*)^{k\theta} \over
\Gamma((k+1)\theta)}$ is a series representation of the special Fox function
[{\it Gl\"ockle and Nonnenmacher}, 1993] (see the Appendix for details).

The ETAS model has been simulated numerically using the algorithm
described in [{\it Ogata}, 1998, 1999].
Starting with a large event of magnitude $M$ at time $t=0$, events are
then simulated sequentially. After each event, we calculate the conditional
intensity $\lambda(t)$ defined by
$$
\lambda(t)=\sum_{t_i\leq t} {K~10^{\alpha(m_i-m_0)} \over (t-t_i+c)^{1+\theta}}
$$
where $t$ is the time of the last event and $t_i$ and $m_i$ are the 
times and magnitudes
  of all preceding events that occurred at time $t_i\leq t$.
The time of the following event is then determined according to the
non-stationary Poisson process of conditional intensity $\lambda(t)$,
and its magnitude is chosen in a Gutenberg-Richter distribution with
parameter $b$.
Theses simulations are compared to the theoretical
predictions in Figure \ref{fig2ninf1}, which shows the
aftershock seismic rate $\lambda(t)$ in the sub-critical regime triggered by a
main event of $M= 6.8$, for the parameters
$K= 0.024$ (constant in (\ref{first})), the threshold $m_0= 0$ for
aftershock triggering,
$c= 0.001$, $\alpha= 0.5$, a $b$-value $b= 1.0$ and $\theta= 0.2$ (corresponding
to a local Omori's exponent $p=1.2$). These parameters lead
to a branching ratio $n=0.95$ (equation (\ref{second}))
and a characteristic cross-over time $t^*=4500$ (equation (\ref{ngnvl})).
The noisy black line represents the seismicity rate obtained for the
synthetic catalog.
The local Omori law with exponent $p=1+\theta=1.2$ is shown for reference as
the dotted line. The analytical solution (\ref{bhgknaalkal2}) is
shown as the thick line.
The two dashed lines represent the approximation solutions (\ref{eadfgz1})
for $t<t^*$ and (\ref{eaz1}) for $t>t^*$.

\subsection{The super-critical regime $n>1$ and $\theta>0$}

  From the definition of the branching ETAS model for $n>1$, it is
clear that the number of events  $\lambda(t)$ blows up exponentially for large times
as $n-1$ to a power proportional to the number $t$ of generations. We
shall show below that the rate of the exponential growth can be
calculated explicitly, which yields $\lambda(t) \sim e^{t/t^*}$, where
$t^*$ has been defined in (\ref{ngnvl}).
However, there is
an interesting early and intermediate time regime in the situation where
a great event of magnitude $M$
has just occurred at $t=0$. In this case, the total seismicity is the
result of two competing effects: (1) the total
seismicity tends to decay according to the Omori's law
governing the rate of daughter-earthquakes triggered by the great
event; (2) since
each daughter may in turn trigger grand-daughters, grand-daughters may
trigger grand-grand daughters
and so on with a number $n>1$ of children per parent, the induced
seismicity will eventually
blow up exponentially. However, before blowing up, one can expect
that seismicity
will first decay because it is mainly controlled by the large rate
$\sim 10^{\alpha (M-m_0)}$
directly induced by the great earthquake which decays according to its
``local'' Omori's law. This decay will be progressively perturbed
by the proliferation of daughters of daughters of ... and will
cross-over to the
explosive exponential regime.

At early times $c \ll t \ll t^*$, the early decay rate of
aftershocks is the same $\approx 
(S_0 / \Gamma(\theta) (n-1))~({t^*}^{-\theta} /t^{1 - \theta})$ as for the
sub-critical regime (\ref{eadfgz1}) (see the Appendix).
However, as time increases, the Appendix shows that the decay of
aftershock activity can be represented as a power law with an
effective apparent exponent $\theta_{\rm app}>\theta$
increasing progressively with time.
The seismic rate will thus decay approximately as $\sim
1/t^{1 - \theta_{\rm app}(t)}$.
Quantitatively, the large time behavior is (see the Appendix)
\be
\lambda(t) \sim {S_0 \over (n-1) t^* \theta}~~ e^{t/t^*}  \label{gnalksd1}
\ee
exhibiting an exponential growth at large times. Expression (\ref{ngnvl})
shows that $1/t^* \sim |1-n|^{1 \over \theta}$. Thus, as expected,
the exponential growth disappears as $n \to 1^+$.

The full expression of $\lambda(t)$ valid at  times $t \gg c$ is
\be
\lambda(t) = {S_0 \over (n-1) }~{{t^*}^{-\theta} \over t^{1-\theta}}~
\sum_{k=0}^\infty {(t/t^*)^{k\theta} \over \Gamma((k+1)\theta)}
\label{bhgknaalkalx}
\ee
Expression (\ref{bhgknaalkalx}) provides the solution that describes
the cross-over from the $1/t^{1-\theta}$ Omori's law at early times
(\ref{eadfgz1}) to the exponential growth (\ref{gnalksd1}) at large times.


Figure \ref{fig2sup1ll} tests these predictions by comparing them with
direct numerical simulation of the ETAS model, in the case of a main shock
of magnitude $M= 6$. The parameters of the synthetic catalog are
$K= 0.024$ (constant in (\ref{first})), the threshold $m_0= 0$ for
aftershock triggering,
$c= 0.001$, $\alpha= 0.5$, a $b$-value $b= 0.75$ and $\theta= 0.2$ (corresponding
to a local Omori's exponent $p=1.2$). These parameters lead
to a branching ratio $n=1.43$ (equation (\ref{second}))
and a characteristic cross-over time $t^*=0.85$ (equation (\ref{ngnvl})).
The noisy black line represents the seismicity rate obtained for the
synthetic catalog.
The local Omori law with exponent $p=1+\theta=1.2$ is shown for reference as
the dotted line.
The analytical solution (\ref{bhgknaalkalx}) is shown as the thick line.
The two dashed lines correspond to the approximative analytical solutions
(\ref{eadfgz1}) and (\ref{gnalksd1}). At early times $c<t<t^*$,
the decay of $N(t)$
is initially close to the prediction (\ref{eadfgz1}).
For $t>t^*$, we observe that
the analytical equation (\ref{bhgknaalkalx}) is very close to the exponential
solution (\ref{gnalksd1}).

\subsection{Case $\theta <0$ corresponding to a local Omori's law
exponent $p<1$}

We have already remarked that, in this case,
the integral $\int_0^\infty {dt \over (t+1)^{1+\theta}}$ in the definition (\ref{second})
of the branching ratio $n$ becomes unbounded:
the number of daughters created beyond any time $t$ far
exceeds the number of daughters created up to time $t$.

The appendix shows that the general equation (\ref{third}) still
holds and the general
derivation starting with (\ref{fourth}) up to (\ref{bghnglal}) still applies.

Similarly to the super-critical case $n>1$ of the regime $\theta >0$,
 we find a crossover from a power-law decay at early times to an
 exponential increase of the seismicity rate at large times.
The characteristic time $\tau$ that marks the transition between these two regimes is given by
\be
\tau=c \left( {n_0 \Gamma(|\theta|) \over 1 + {n_0 \over
|\theta|}}\right)^{- {1 \over |\theta|}}~.
\label{jgjjgswsa}
\ee

In contrast with the case $\theta >0$, the early time behavior ({\it i.e.}, $c \ll t \ll \tau$)
 of the global decay law in the case $\theta <0$ is similar to the local Omori law :
\be
\lambda(t)= {S_0 \over (1 + {n_0 \over|\theta|}) \Gamma(|\theta|)}
 ~ { \tau ^{-|\theta|} \over t^{1-|\theta|}}
\label{gnlanvlaax2}
\ee

Similarly to the super-critical case $n>1$ of the regime $\theta >0$, the
long time dependence of the regime $\theta <0$ is controlled by a simple pole
$1/\tau$ leading to a long-time seismicity growing
exponentially
\be
\lambda(t) = {S_0 \over (1 + { n_0 \over |\theta|}) \tau |\theta|}  ~ e^{t/\tau}
\label{expogjjgd}
\ee
This result is in agreement with the fact that
the number of daughters born up to time $t$
is an unbounded increasing function of $t$, and we should thus recover
a regime similar to the super-critical case of $\theta >0$.

The full expression of $\lambda(t)$ valid at times $t>c$ is
\be
\lambda(t) = {S_0 \over (1 + { n_0 \over |\theta|}) }~{1 \over t}~
\sum_{k=1}^\infty {(t/\tau)^{k|\theta|} \over \Gamma(k|\theta|)}
\label{bhgknaalkal0}
\ee
Expression (\ref{bhgknaalkal0}) provides the solution that describes
the cross-over from the local Omori law $1/t^{1-|\theta|}$ at early times to
the exponential growth at large times.

Figure \ref{thetainf0} compares these predictions to a
direct numerical simulation of the ETAS model, in the case of a main shock
of magnitude $M= 7$. The parameters of the synthetic catalog are
$K= 0.02$, $m_0= 0$,
$c= 0.01$, $\alpha= 0.5$, $b= 1$ and $\theta=-0.1$ (corresponding
to a local Omori's exponent $p=0.9$). These parameters lead
to a characteristic cross-over time $\tau=10^5$ (equation (\ref{jgjjgswsa})).
The noisy black line represents the seismicity rate obtained for the
synthetic catalog.
The local Omori law with exponent $p=1+\theta=0.9$ is shown for reference as
the dotted line.
The analytical solution (\ref{bhgknaalkal0}) is shown as the thick line.
The two dashed lines correspond to the approximative analytical solutions
(\ref{gnlanvlaax2}) and (\ref{expogjjgd}). At early times $c<t<\tau$,
the decay of $\lambda(t)$ is initially close to the prediction (\ref{gnlanvlaax2}).
For $t>\tau$, we observe that the analytical equation (\ref{bhgknaalkal0})
 is very close to the exponential solution (\ref{expogjjgd}).

\section{Discussion}

Assuming that each event triggers aftershock sequences according to the local
   Omori law with exponent $1+\theta$, we have shown that the decay law
of the global aftershock sequence is different from the local one.
Depending on the branching ratio $n$, which is a function of all ETAS
parameters,
we find two different regimes, the sub-critical regime for $n<1$ and
the super-critical regime for $n>1$ and $\theta>0$. For the two regimes
in the case $\theta>0$, a characteristic time $t^*$,
function of $c$, $n$ and $\theta$, appears in the global decay law $\lambda(t)$ and
marks the transition between the early time behavior and the large
time behavior.
In the sub-critical regime ($n<1$), the global decay law is composed
of two power laws.
At early times ($t<t^*$), $\lambda(t)$ decays like $t^{-1+\theta}$. At large times
   ($t>t^*$) the global decay law recovers the local law $ N(t) \sim
t^{-1-\theta}$.
In the super-critical regime ($n>1$ and $\theta>0$), the early times  decay 
law is similar to that of the sub-critical regime, and the seismicity
 rate increases exponentially for large times.
The case $\theta<0$ leads to an infinite $n$-value, due to the slow
 decay with time of the local Omori law. In this case, 
we find a transition from an Omori law with exponent
$1-|\theta|$ similar to the local law, to an exponential increase at large times, 
with a crossover time $\tau$ different from the characteristic 
time $t^*$ found in the case $\theta>0$.
 Thus, the Omori law is only an approximation of the global
   decay law valid for some time periods and parameter values.
The value of the local Omori exponent $p=1$ is the only one for which the local
and the global decay rate are similar, and are both power-laws
without any characteristic time.
For small $n$, $t^*$ is very small so that in real data we should
observe only the
behavior $t>t^*$ characteristic of large times. The global
decay law then appears similar to the local Omori law.
On the contrary, for $n$ close to 1, $t^*$ is very large by comparison with
the time period available in real data, and we should observe only the
power-law behavior $ \lambda(t) \sim t^{-1+\theta}$ characteristic of early times,
with a global $p$-value smaller than the local one. Changing $n$ thus provides
an important source of variability of the exponent $p$.

\subsection{Estimation of $n$ and $t^*$ in earthquake data}

In real earthquake data, it is possible to evaluate the branching value $n$
in order to determine if the seismic activity is either in the sub-
or the super-critical regime.
The values of $n$ and $t^*$ can be evaluated from equations
(\ref{second}) and (\ref{ngnvl})
as a function of the ETAS parameters  $b$, $p=1+\theta$, $c$, $K$ and $\alpha$.
The parameters of the ETAS model and their standard error can be inverted from
seismicity data (time and magnitudes of each event) using a maximum 
likelihood method [{\it Ogata}, 1988].
We now discuss the range of the different parameters obtained from
such inversion procedure.

\begin{itemize} 
\item The parameter $\alpha$ is  found to vary between $0.35$ to
$1.7$, and is often close to
$0.5$ [{\it Ogata}, 1989, 1992; {\it Guo and Ogata}, 1997].
An $\alpha$-value of $0.5$ means that a mainshock of magnitude $M$
will have on
average 10 times more aftershocks than a mainshock of magnitude
$M-2$, independently of $M$.
Note that our definition of $\alpha$ is slightly different from that used by
{\it Ogata} and we have divided his $\alpha$-values by $\ln(10)$ to 
compare with our definition.

For some seismicity sequences, {\it Ogata} [1989, 1992] and {\it Guo and
Ogata} [1997]
found $\alpha>b$.
According to (\ref{second}), this leads to an infinite $n$-value
if we use a Gutenberg-Richter magnitude distribution.
As we said, a truncation of the magnitude distribution is needed
to obtain a physically meaningful finite $n$-value because the seismicity rate
is controlled by the largest events.

A large $\alpha$-value can be associated with seismic activity
called ``swarms'', while a small $\alpha$-value is observed for
aftershock sequences
with a single mainshock and no significant secondary aftershock sequences
[{\it Ogata}, 1992, 2001].

\item The parameter $c$ is usually found to be
of the order of one hour [{\it Utsu et al.}, 1995].
In practice, the evaluation of $c$ is hindered by the incompleteness
of earthquake catalogs just after the occurrence of the mainshock,
due to overlapping aftershocks on the seismograms.  A large $c$
is often an artifact of a change of the
detection threshold. Notwithstanding these limitations,
well-determined non-zero $c$-value have been
obtained for some aftershocks sequences [{\it Utsu et al.}, 1995]. Note that
a non-zero $c$ is required for the aftershocks rate to be finite just
at the time of the mainshock.

\item The ``local'' $p$-value, equal to $1+\theta$, describes the decay 
law of the aftershock sequence triggered by a single earthquake.
The local Omori law is the law $\phi(t)$ obtained by inverting 
the ETAS model on the data.
The ``global'' $p$-value describes the decay law of the whole aftershock
sequence, composed of all secondary aftershocks triggered by each aftershock.
 The global Omori law is the law $\phi(t)$ fitted directly on the data,
 without taking into account the hierarchical structure of branching of the ETAS model.
We have shown that the Omori law is only an approximation of the 
global decay law, so that in the subcritical regime the global $p$-value
 will change from $1-\theta$ at early times to $1+\theta$ at large times.
[{\it Guo and Ogata}, 1997] measured both the local and global $p$-values
for 34 aftershocks sequences in Japan, and found that
the local $p$-value is usually slightly larger than the global $p$-value
[{\it Guo and Ogata}, 1997].
This is in agreement with our prediction
when identifying the local $p$-value with  $1+\theta$ (recovered at
large times) and the global $p$-value with $1-\theta$ found at early times.
{\it Guo and Ogata} [1997] and {\it Ogata} [1992, 1998, 2001]
found a local $p$-value smaller than one for some aftershocks
sequences in Japan.
Within the confine of the ETAS model, this corresponds to the case
$\theta <0$ discussed above and in the appendix.

\item The parameter $K$ measures the rate of aftershocks triggered by
each earthquake,
independently of its magnitude. Recall that the
branching ratio $n$ is proportional to $K$. It is usually found
of the order of $K \approx 0.02$ [{\it Ogata}, 1989, 1992;
   {\it Guo and Ogata}, 1997], but large variations of $K$-value from 0.001
to 5 are reported by {\it Ogata} [1992].

\item The parameter $\mu$ measures the background seismicity rate
that is supposed
to arise from the tectonic loading. $\mu \simeq 0$ for an aftershock sequence
triggered by a single mainshock. This parameter as no influence on the
branching ratio $n$. In real catalogs, the background seismicity only accounts
for a small part of the seismic activity.

\end{itemize}

We have computed the branching ratio $n$ and the cross-over time
$t^*$ from the ETAS parameters measured by {\it Ogata} [1989, 1992]
 for several seismicity sequences in Japan and elsewhere.
 The ETAS parameters and the $n$ and $t^*$ values are given
 in Table \ref{tab1}. When the $b$-value is not given in the text,
we have computed $n$ and $t^*$ assuming a $b$-value is equal to $1$.
We find that the $n$-value is either smaller or larger than $1$. 
This means that the seismicity can be interpreted to be either in
 the sub- or in the super-critical regime.
An infinite $n$-value is found if the local  $p$-value is smaller than one ($\theta<0$) or if the
$\alpha$-value is larger than the $b$-value.
 For the same area, the ETAS parameters and the $n$ and $t^*$ values are found to vary in time, 
sometimes changing from the sub- to the super-critical regime.
The characteristic time $t^*$ shows large spatial and temporal variability,
 ranging from 0.4 days to $10^{22}$ days. Large $t^*$ values are related to
 a branching ratio $n$ close to one, {\it i.e.}, close to the critical point $n=1$. 
The ETAS model thus provides a picture of seismicity in which sub-critical and
super-critical regimes are alternating in an intermittent fashion. As we
shall argue, the determination of the regime may provide important clues
and quantitative tools for prediction.

\subsection{Implications of the ETAS model in the sub-critical regime $n<1$}

In the sub-critical regime, the ETAS model can explain many of the
departures of the
global aftershock decay law from a pure Omori law.

The ETAS model contains by definition (and thus ``explains'')
the secondary aftershock sequences
triggered by the largest aftershocks that are often observed
[{\it Correig et al.}, 1997; {\it Guo and Ogata}, 1997;  {\it
Simeonova and Solakov}, 1999; {\it Ogata}, 2001].
In the ETAS model, the fact that secondary aftershock sequences of
large aftershocks
can stand out above the overall background aftershock seismicity 
results from the
factor $10^{\alpha (m_i-m_0)}$ in (\ref{first}).

Our analytical results may rationalize why some alternative models
of aftershock decay work better than the simple modified Omori law.
In the sub-critical regime, we predict an increase of the apparent
global $p$-value from $1-\theta$ at early times to $1+\theta$ at large times.
To our knowledge, this change of exponent has never been observed.
This change of power law may be approximated by the stretched
exponential function proposed by
[{\it Kissinger}, 1993; {\it Gross and Kisslinger}, 1994] to fit
aftershocks sequences.
In the stretched exponential model, the rate of aftershocks $\lambda(t)$ is defined by
\be
	\lambda(t)=K~t^{q-1}e^{-(t/t_0)^q}~,
\ee
where  $q$, $K$ and $t_0$ are constants.
At early times, this function decays as a power law $1/t^{1-q}$ with
apparent Omori's exponent $1-q$. For times
larger than the relaxation time $t_0$, the seismicity rate decays exponentially
in the argument $(t/t_0)^q$. For $q<1$, this decay is much slower than
exponential and can be accounted for by an apparent power law with
larger exponent.
Figure \ref{stretched} compares the stretched exponential function
  with the analytical solution of the ETAS model (\ref{bhgknaalkal2})
with parameters
$t^*=t_0$ and $\theta=q$, and with the Omori law of exponent $p=1-q$.
These three laws have the same power-law behavior at early times, and then
both the stretched exponential and the analytical solution (\ref{bhgknaalkal2})
decay faster than the Omori law at large times.
The fact that it is very difficult to distinguish the decay laws described
by power laws and by stretched exponential has been illustrated in
[{\it Laherr\`ere and Sornette}, 1998] in many examples including earthquake
size and fault length distributions.
{\it Kissinger} [1993] and {\it Gross and Kisslinger}, 1994] compared
this function to the modified Omori law $\lambda(t)=K~(t+c)^{-p}$ for several
aftershocks sequences in  southern California.
They found that the stretched exponential fit often works better for
the sequences with a small $p$-value or a large $q$-value, indicative
of a slow decay for small times.
This is in agreement with our result that in the sub-critical regime a slowly
decaying aftershock sequence (global $p$-value smaller than one) will
then cross-over to a more rapid decay for time larger than $t^*$.
The relaxation time $t_0$ ranges between
2 days and 380 days for the sequences that are better fitted by the stretched
exponential [{\it Kissinger}, 1993]. This parameter is analogous to $t^*$
found in our model, because these two parameters define the transition
from the early time power-law decay to another faster decaying behavior
  for large times.
To further validate our results, these aftershocks sequences should be fitted
using equation (\ref{bhgknaalkal2}) to compare our results with the
stretched exponential function and determine if the transformation of the early
time power law decay is better fitted by a stretched exponential
fall-off or an increase
in the apparent Omori exponent from $1-\theta$ to $1+\theta$ as
predicted by our results.

The ETAS model can also rationalize some correlations
found empirically between seismicity parameters. 
It may explain the rather large variability of the global empirical $p$-value.
{\it Guo and Ogata} [1995] have reported a positive correlation between the
   Gutenberg-Richter $b$-value and the $p$-value (exponent of the
global Omori law) for several aftershocks sequences in Japan. A similar correlation has also been
   found by [{\it Kisslinger and Jones}, 1991] for several aftershock sequences
   in southern California, but this correlation was detectable only if
the earthquake sequences were separated into thrust and strike slip events.
This positive correlation between $b$ and global $p$ values is expected from
our analysis.
  From equation (\ref{second}), we see that a small $b$-value is associated
with a large $n$ value. For $n \simeq 1$, the characteristic time
$t^*$ is very large, so that the global aftershock rate decays as a power law with
exponent $1-\theta$ over a large time interval.
For $n>1$ and $\theta>0$, we see an apparent global $p$-value smaller than $1-\theta$
which decreases with time.
In contrast, for large $b$-values, the branching ratio $n$ is small
and the characteristic time $t^*$ is very small. In this case, only the large 
time behavior is observed with a larger exponent $1+\theta$. Consequently,
 in the subcritical regime, our results predict a change of the
global $p$-value from $1-\theta$ for small $b$-value and times $t \ll t^*$  to
$1+\theta$ for large $b$-values.
There is also a positive correlation
between $p$-value and $b$-value in the super-critical regime.
For $n>1$ or $\theta<0$, the global aftershock sequence is characterized by
an apparent exponent $p$ smaller than $1-|\theta|$ which decreases with time.
Then, we expect the apparent exponent $p$ to be all
the smaller, the smaller is the $b$-value, because the characteristic 
times $t^*$ for $\theta>0$ or $\tau $ for $\theta<0$ decreases with $b$.
The variability of the global $p$ exponent
reported by {\it Guo and Ogata} [1995] and {\it Kisslinger and Jones} [1991]
may thus be explained by a change of $b$-value and a constant local
$p$ exponent.
However, the results of {\it Guo and Ogata} [1997] contradict this
interpretation. {\it Guo and Ogata} [1997] studied the same aftershocks sequences than
{\it Guo and Ogata} [1995] but they measured the local $p$-value of
the ETAS model.
They still found a large variability in the local $p$-value, and a positive
correlation between this local $p$-value and the $b$-value.


\subsection{Implications of the ETAS model in the super-critical regime
and in the case $\theta<0$}

In the regime where the mean number of aftershocks per mainshock is larger
than one ({\it i.e.}, $n>1$), the mean rate of aftershocks increases
exponentially for large times. However, because of the statistical fluctuations,
the aftershock sequence has a finite probability to die. This probability of extinction
can be evaluated for the simple branching model without time dependence
[{\it Harris}, 1963]. Therefore, a branching ratio larger than 1 does not imply
necessarily
that the number of aftershocks will be infinite. If $n$ is not too large, and if
the number of aftershocks is small, there is a significant probability that
the aftershock sequences will die, as observed in numerical simulations of
the ETAS model. If the characteristic time $t^*$
is very large, the aftershock sequence may not remain supercritical long enough
for the exponential increase to be observed.
Even if the large times exponential acceleration is rarely observed in real seismicity,
  it may explain the acceleration of the deformation before material failure.
  The early times behavior of the seismic activity preceding the
    exponential increase has also important possible implications
for earthquake prediction, and can rationalize some empirically
proposed seismic  precursors, such as the low $p$-value [{\it Liu}, 1984;
{\it Bowman}, 1997], or the relative seismic quiescence preceding large
aftershocks [{\it Matsu'ura}, 1986; {\it Drakatos}, 2000].

It is widely accepted that about a third to a half of strong
earthquakes are preceded by foreshocks  [e.g., {\it Jones and Molnar}, 1979;
{\it Bowman and Kisslinger}, 1984; {\it Reasenberg}, 1985, 1999;
  {\it Reasenberg and Jones}, 1989; {\it Abercombie and Mori}, 1996],
{\it i.e.}, are preceded by an unusual high seismicity rate for time periods of
  the order of days to years, and distance up to hundreds kilometers.
However, there is no reliable method for distinguishing foreshocks
from aftershocks.
Indeed, the ETAS model makes no arbitrary distinctions between
foreshocks, mainshocks
and aftershocks and describes all earthquakes with the same laws.
While this seems a priori paradoxical, our analysis of the ETAS model
provides a useful tool for identifying foreshocks,
{\it i.e.}, earthquakes that are likely to be followed by a larger event, from usual
aftershocks that are seldom followed by a larger earthquake. The
characterization
of foreshocks will be performed in statistical terms rather than on a
single-event
basis. In other words, we will not be able to say whether any specific event
is a precursor. It is the ensemble statistics that may betray a
foreshock structure.
 
The crux of the method is that, when seismicity falls in the regime
with a branching ratio
$n>1$, the corresponding earthquake sequences
can be identified as foreshocks. This is because the super-critical regime
corresponds to an exponentially accelerating seismicity for
times larger than $t^*$:  by a pure statistical effect,
the larger number of earthquakes of any size will sample more and more
the branch of the Gutenberg-Richter law toward large events. Thus by the sheer
weight of numbers,
larger and larger earthquakes will occur as time increases. Of course, we are
not implying any precise deterministic growth law, but statistically, the
largest events should indeed grow significantly, the more so, the
more within the
super-critical regime, the larger the branching ratio $n>1$.
Conversely, this argument implies that,
in the subcritical regime, the triggered events are usual
aftershocks, because a mainshock is
unlikely to be followed by a larger triggered event.
Foreshock sequences can thus be identified by evaluating the
branching ratio $n$ from the inversion of seismic data (times and magnitudes of an
earthquake sequence) for the ETAS parameters.
There is however a finite probability than a triggered event in the
 subcritical regime be larger than the triggering event, and thus the
 triggering event will be
a foreshock of the triggered event. Therefore, foreshocks
can be observed even in the sub-critical regime, but they are less
frequent than aftershocks.

A note of caution is in order:
the direct estimation of $n$ and $t^*$ or $\tau$ may be quite imprecise if
 the number of events is small. Based on our analysis and our results,
the foreshock regime can be nevertheless identified with relatively
 good confidence if one assumes an upper bound for the local exponent $p$.
Let us assume for instance that
the local $p$-value is smaller than $1.3$ ({\it i.e.}, $\theta<0.3$); according to
our results, the global exponent $p=1+\theta$ cannot become smaller
than $1-\theta=0.7$
in the sub-critical regime. In contrast, in the supercritical regime, we have shown
that the apparent exponent is smaller than or at most equal to $1-\theta$.
Therefore, a measure of the
global $p$-value yielding a value smaller than $0.7$, is always associated
with the super-critical regime.
As we said above, {\it Guo and Ogata} [1997] and {\it Ogata} [1992, 1998, 2001]
found a local $p$-value smaller than one for some aftershocks
sequences in Japan corresponding to the case $\theta <0$.
A small global $p$-value can thus also result
from a small local $p$-value. In sum, a small global $p$-value
results either from a larger than one local $p$-value in the supercritical
regime $n>1$ or from a small (smaller than $1$) local $p$-value before
the exponential growth regime.

\begin{figure}


\psfig{file=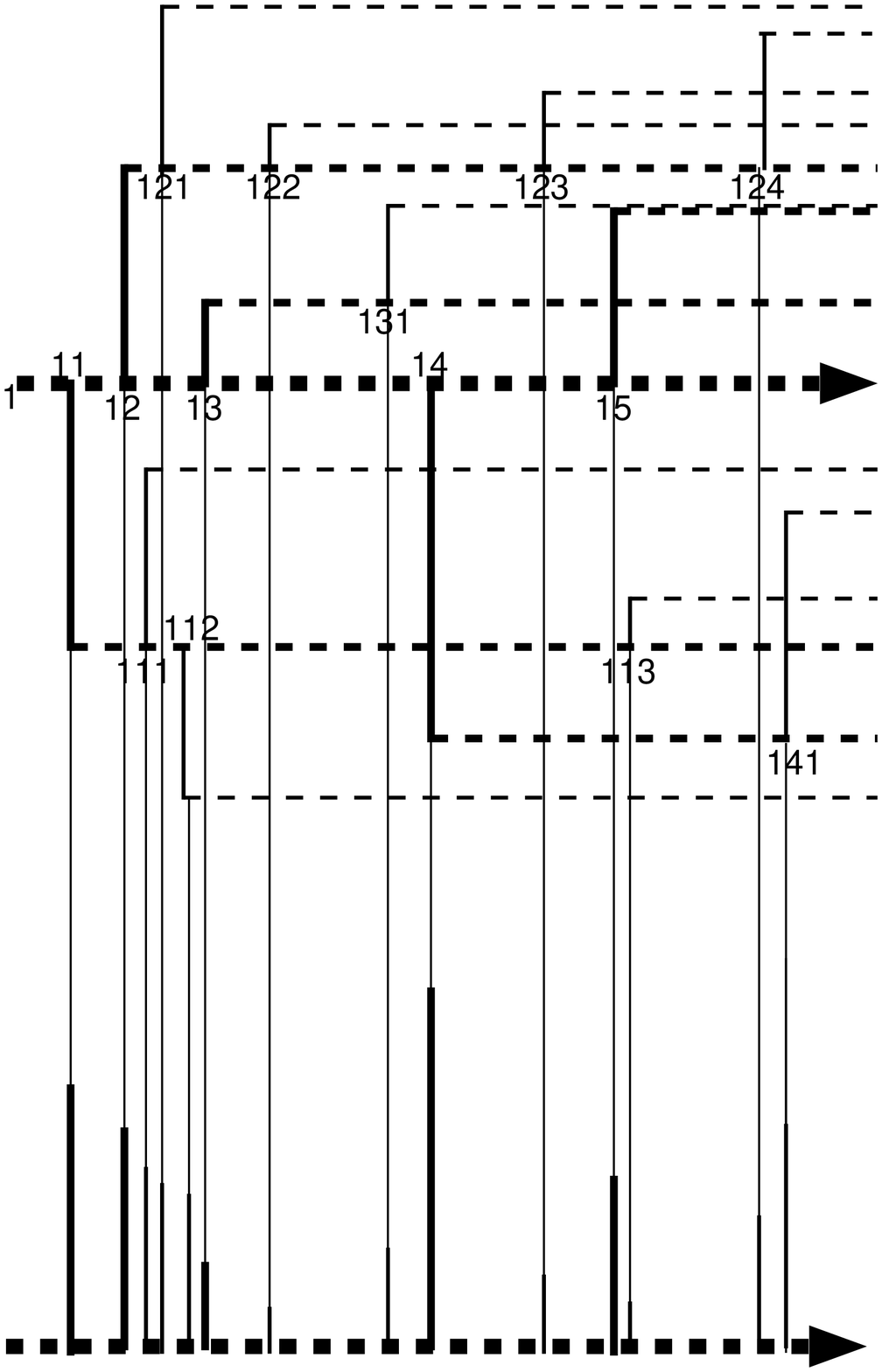,width=8cm}
\caption{\label{fig1branch} Schematic representation of the branching process
associated with the ETAS model defined by (\ref{first}) and (\ref{third}).
In this example, the thickest dashed line is the time arrow associated with the
main shock indicated as `$1$'. This main shock triggered five
aftershocks denoted
`$11$', `$12$', `$13$', `$14$' and `$15$' whose magnitudes are proportional to
the length of their vertical lines (their position above or below the thickest
dashed line is arbitrary and chosen to ensure a better visibility of
the diagram).
The aftershock `$11$' triggered three aftershocks denoted `$111$', `$112$' and
`$113$'. The aftershock `$12$' triggered four aftershocks denoted
`$121$', `$122$', `$123$' and `$124$'. The aftershock `$13$' triggered a single
aftershock denoted `$131$'. The aftershock `$14$' also triggered a single
aftershock denoted `$141$'. The aftershock  `$15$' did not trigger
any aftershock.
The observable catalog is the superposition of all these events which are
projected on the thick dashed line at the bottom of the figure, keeping
the thickness as a code for the generation number of each event.
}
\clearpage
\psfig{file=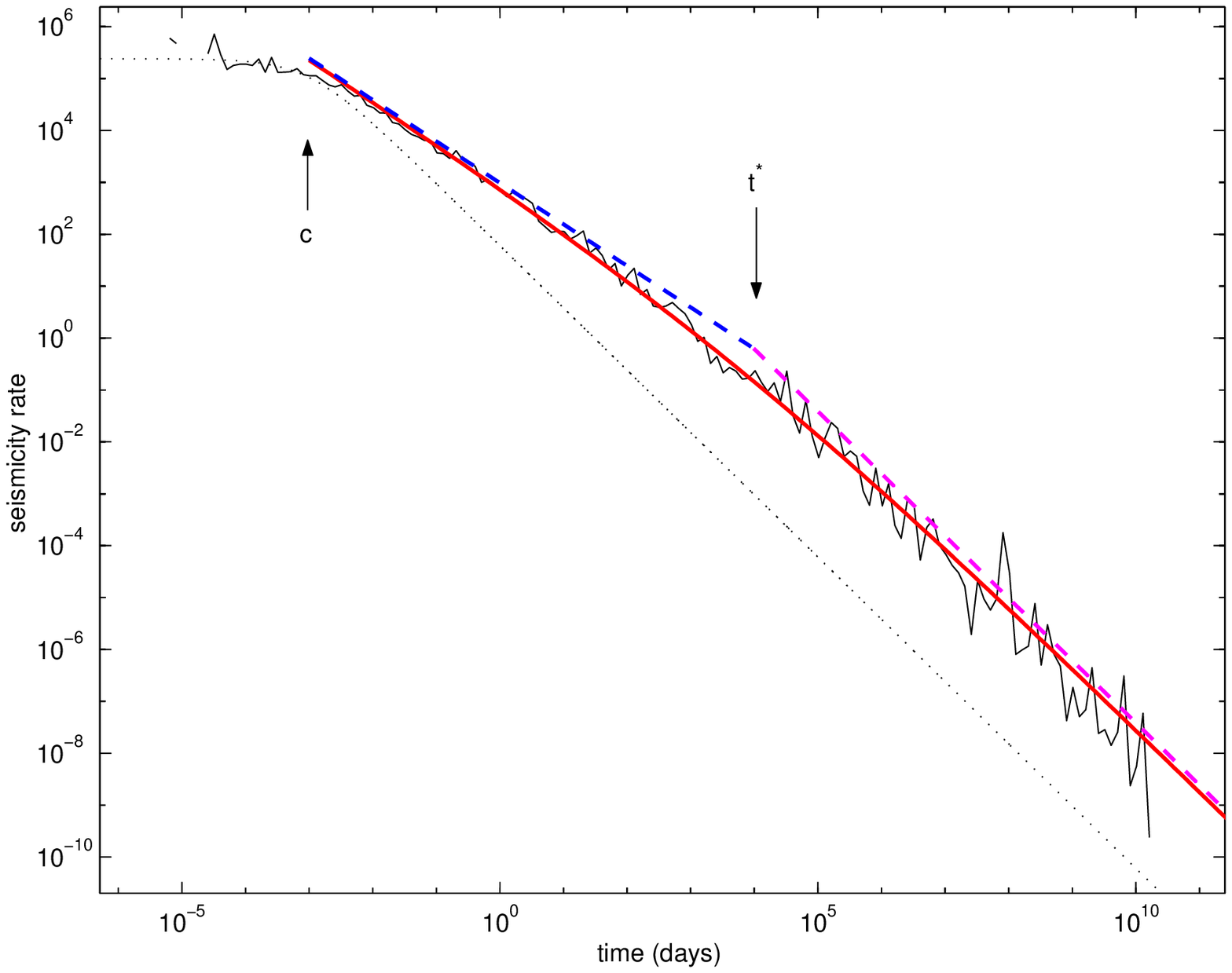,width=16cm}
\caption{\label{fig2ninf1} Seismicity rate $N(t)$ in the sub-critical regime
with $n=0.95$.
The noisy black line represents the seismicity rate obtained for a 
synthetic catalog
generated using $K= 0.024$, $M= 6.8$, $m_0= 0$, $c= 0.001$ day, $\alpha= 0.5$,
$b= 1.0$ and $\theta= 0.2$, giving the characteristic time is $t^*=4500$ days.
The local Omori law with exponent $p=1+\theta=1.2$ is shown for
reference (dotted line).
The analytical solution (\ref{bhgknaalkal2}) is shown as the thick line.
The two dashed lines represents the asymptotic solutions (\ref{eadfgz1})
for $t<t^*$ and (\ref{eaz1}) for $t>t^*$.
}
\clearpage
\psfig{file=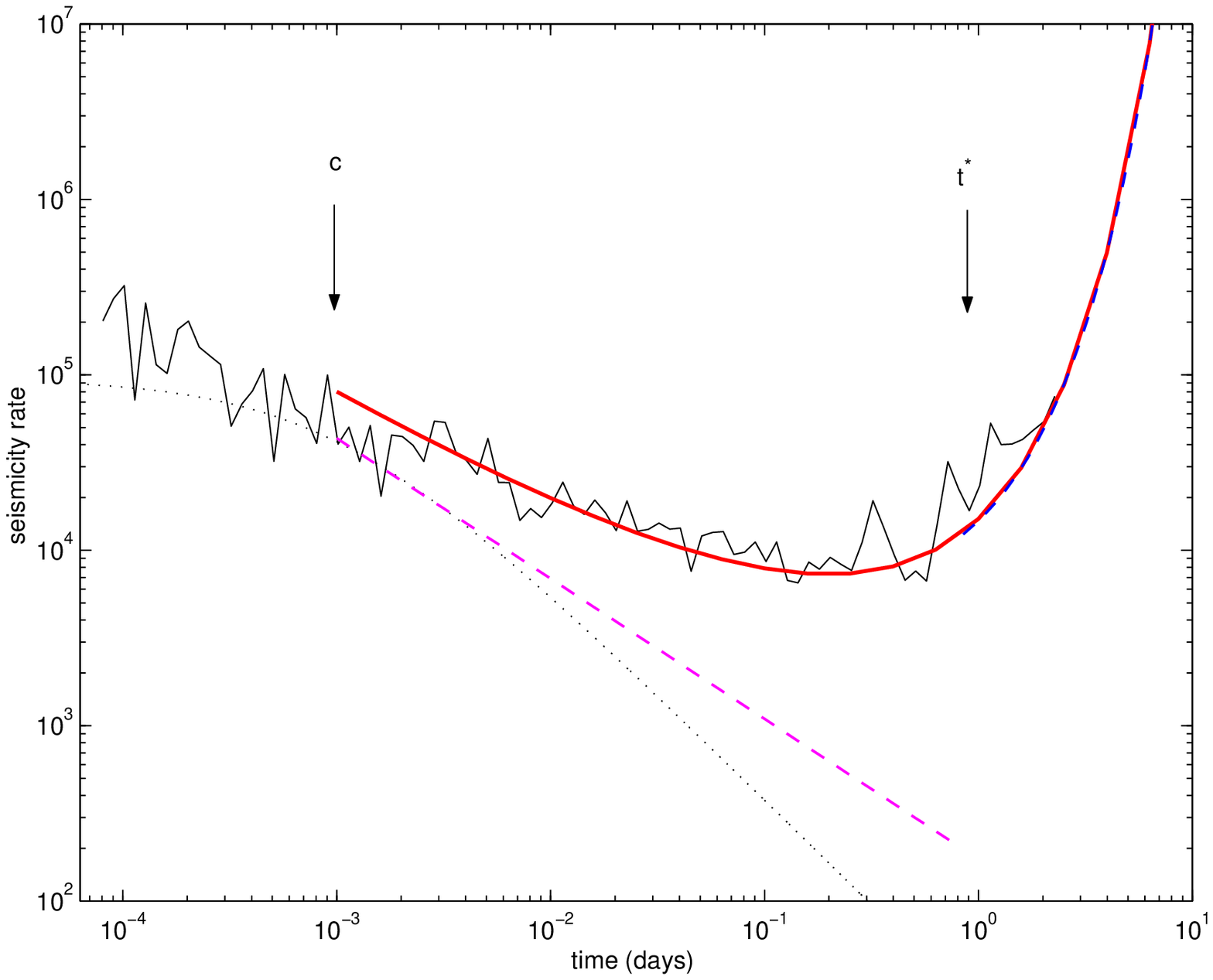,width=16cm}
\caption{\label{fig2sup1ll} Seismicity rate $N(t)$ in the
super-critical regime.
Same legend as in Figure \ref{fig2ninf1}.
The synthetic catalog was generated using the same parameters as for Figure
\ref{fig2ninf1}, except for a lowest $b$-value of $b=0.75$ and a
smallest mainshock
magnitude $M=6$, leading to a branching number $n=1.43$ and a
characteristic time $t^*=0.85$ day.
The analytical solution (thick line) is calculated from equation
(\ref{bhgknaalkalx}).
The two dashed lines correspond to the approximative analytical solutions
(\ref{eadfgz1}) and (\ref{gnalksd1}).}

\clearpage
\psfig{file=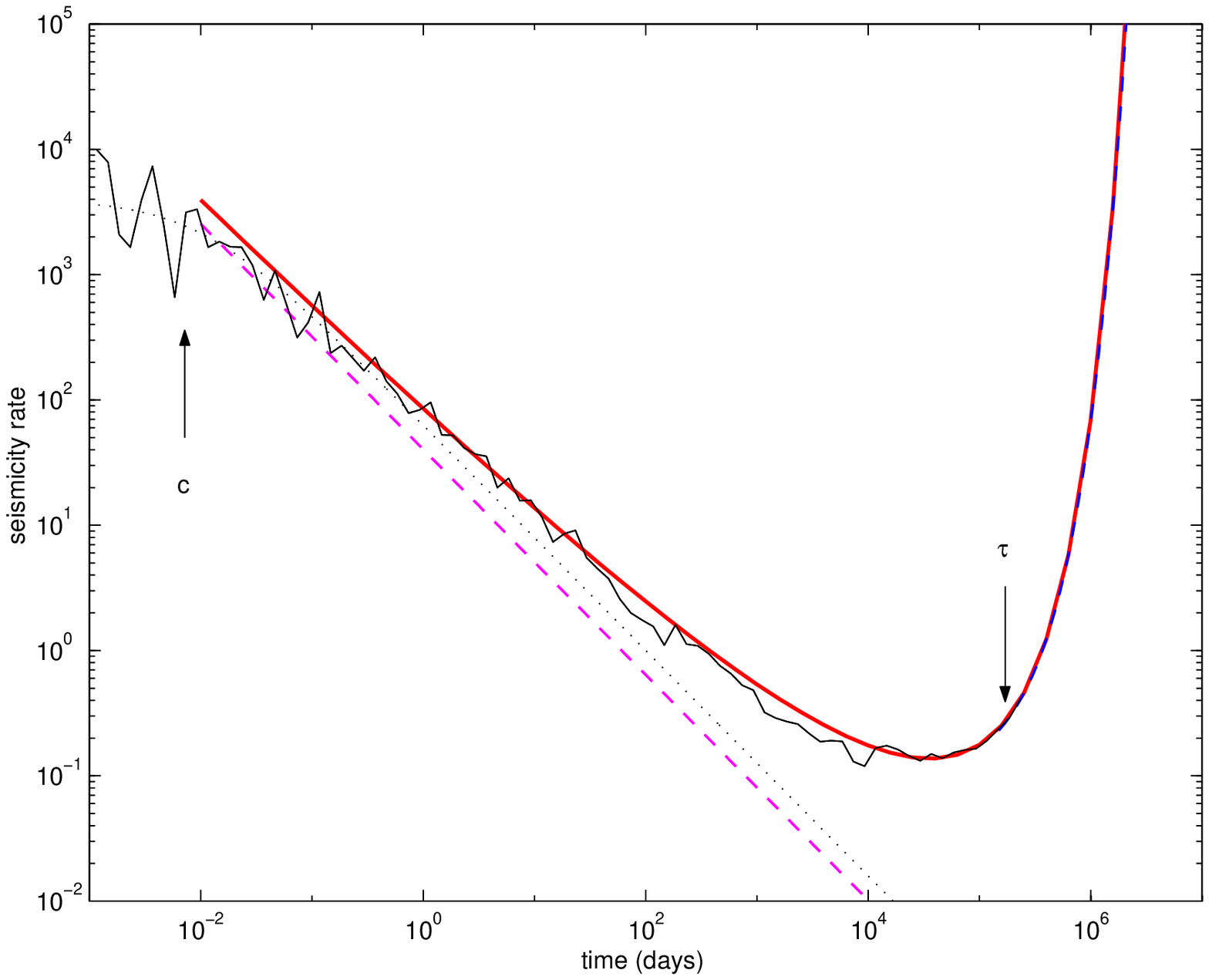,width=16cm}
\caption{\label{thetainf0} Seismicity rate $N(t)$ in the
case $\theta <0$ corresponding to a local Omori's law exponent $p<1$.
Same legend as in Figure \ref{fig2ninf1}.
The synthetic catalog was generated using
$K= 0.02$, $M= 7$, $m_0= 0$, $c= 0.01$ day, $\alpha= 0.5$,
$b= 1.0$ and $\theta= -0.1$, giving the characteristic time is $\tau=10^5$ days.
The analytical solution (thick line) is calculated from equation
(\ref{bhgknaalkal0}).
The two dashed lines correspond to the approximative analytical solutions
(\ref{gnlanvlaax2}) and (\ref{expogjjgd}).}

\clearpage

\psfig{file=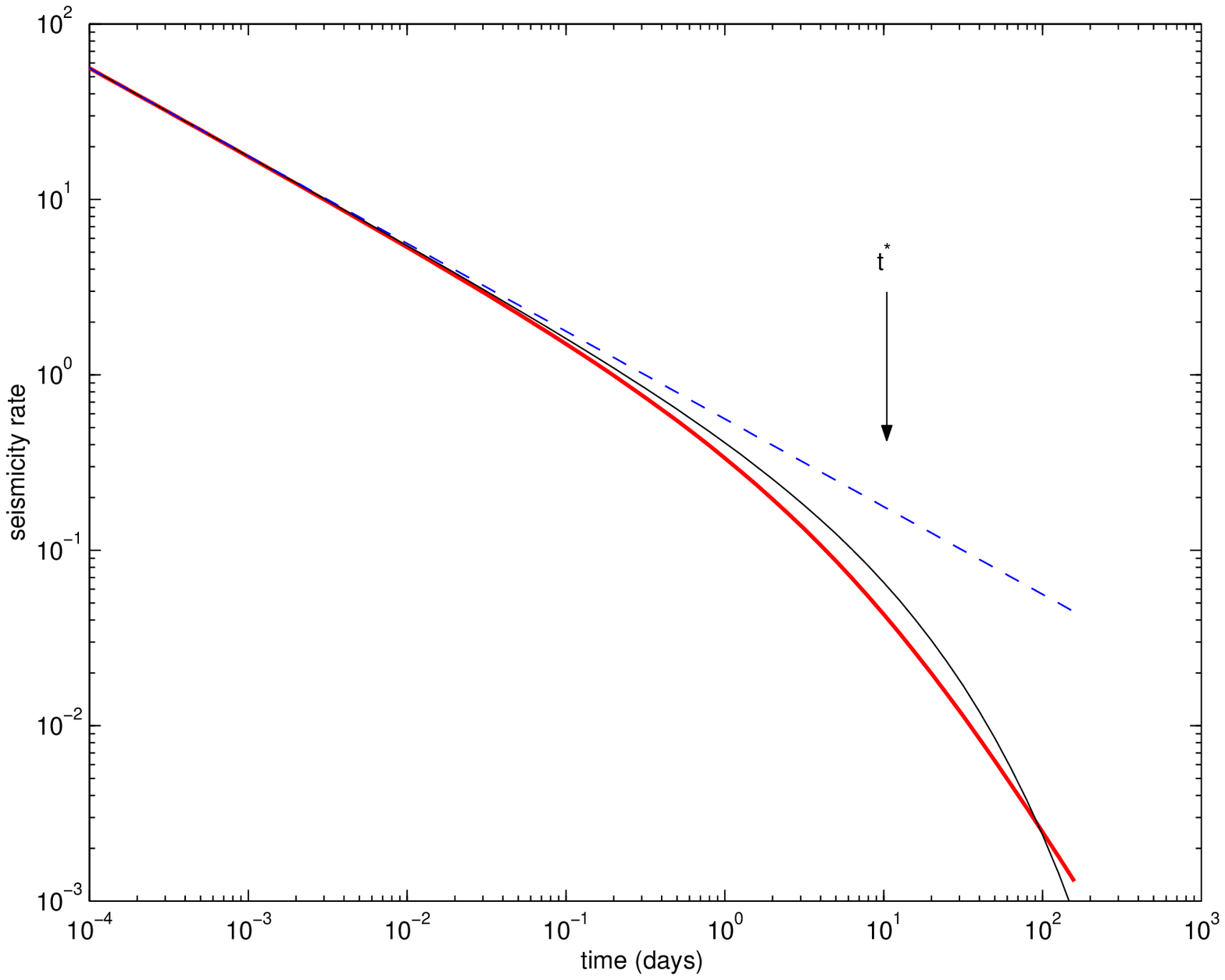,width=16cm}
\caption{\label{stretched} Comparison between the three decay laws of
aftershock sequences:
Omori law with $p=0.7$ (dashed line), stretched exponential with $q=1.3$
and $t_0=10$ days (thin black line) and our analytical solution in the
sub-critical regime
(\ref{bhgknaalkal2}) for $\theta=q=1-p=0.3$ and $t^*=t_0=10$ days (solid gray line).
At early times $t<<t^*$, the three functions are similar and decay as
$t^{-0.7}$.
At large times, the stretched exponential function and the analytical
solution of the ETAS model decay more rapidly that the Omori law.
For times up to $t=10~t^*$, the stretched exponential function is a good
approximation of the ETAS model solution, and describes the transition from
a power law decay at early times to a faster decay law.}

\clearpage

\psfig{file=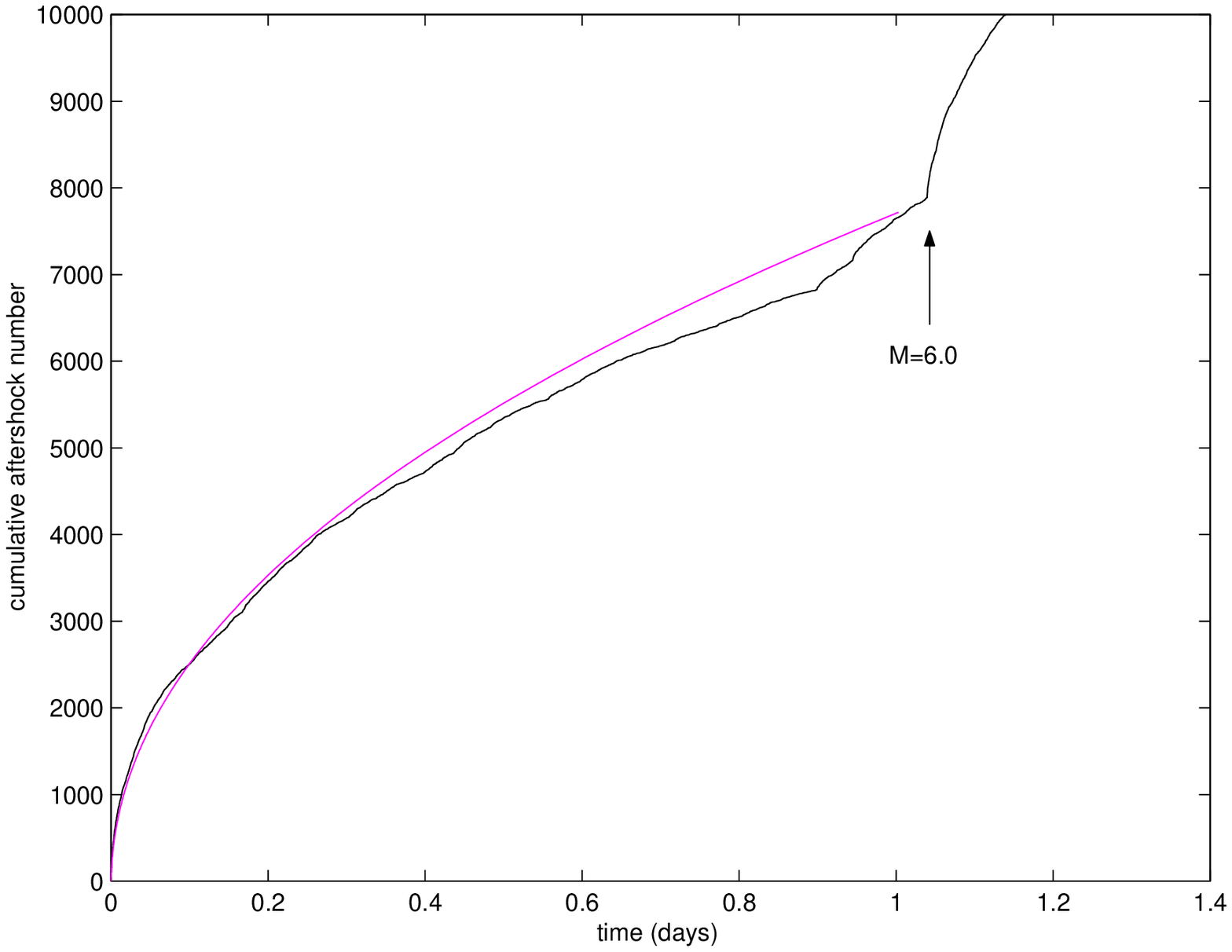,width=16cm}
\caption{\label{quiescence} Cumulative aftershock number in the
   super-critical regime from a synthetic catalog generated using
branching ratio
$n=1.27$, $\theta=0.2$  and $t^*=4.6 $ days. The mainshock magnitude is $M=7.0$.
The thin line is a fit by an Omori law evaluated for time before the occurrence
of the first $M\geq6.0$ aftershock. This fit gives an apparent global
$p$-value of 0.58.
Relative seismic quiescence (by comparison with an Omori law)
is observed before the occurrence of the $M=6.0$ aftershock, due to
the transition
   from an Omori law decay with exponent $p=1-\theta=0.8$ for time $t<<t^*$
   to an exponential increase of the seismicity rate for time $t>>t^*$. }'
\end{figure}

Such a small $p$-value precursor was first proposed empirically by
{\it Liu} [1984], who
studied several aftershocks sequences of moderate earthquakes that
have been followed
by a large earthquake. He proposed that a $p$-value smaller than $1$
is a signature of a
foreshock sequence, whereas $p>1$ is associated with normal aftershock
sequences with a single mainshock in the past. He suggested that
$p$-values close to one characterize
double-mainshock sequences. These empirical rules are part of the earthquake
prediction method used in China [{\it Liu}, 1984; {\it Zhang et al.}, 1999].
The small precursory $p$-value has been used with other precursors to predict
the occurrence of a $M=6.4$ earthquake in China following an other
$M=6.4$ earthquake three months later [{\it Zhang et al.}, 1999].
A precursor associated with a small global $p$-value has
also been observed by {\it Bowman} [1997]
for a sequence in Australia. In 1987, several $M=4-5$ earthquakes occurred
in a region that was not seismically active before, and triggered a
large number of aftershocks characterized by an abnormally low $p$-value of $0.3$.
A sequence of three $M\geq6$ occurred one year later, followed by an
aftershock sequence with a more standard $p$-value of $1.1$.
{\it Simeonova and Solakov} [1999] have also reported a very low
$p$-value of 0.5, for one sequence of aftershocks in Bulgaria, that was followed
one year latter by a larger earthquake. The first part of the aftershock sequence
was well fitted by a modified Omori law, and then a significant deviation occurred
with an abnormally high aftershock rate by comparison with the prior trend.
This departure from an Omori law is expected from our results for an aftershock
sequence in the super-critical regime and the very low value of the exponent $p$
can be interpreted as the apparent exponent within the cross-over from the
$1/t^{1-\theta}$ decay (\ref{eadfgz1}) at early times to the exponential
explosion (\ref{gnalksd1}) at times $t>t^*$ (see Figure \ref{fig2sup1ll}).

In addition to the small precursory $p$-value predicted in the
regime $n>1$, we have shown that this regime is also characterized by a
decrease of the apparent global $p$-value with time.
Such a decrease of $p$-value has also been identified as a precursor
by {\it Liu} [1984].

Other patterns may be a signature of the super-critical regime.
The relative precursory quiescence suggested by {\it Drakatos} [2000]
may also be explained by our results.
In contrast to the ``absolute'' quiescence which detects changes in the
  background seismicity after removing the aftershocks from the catalog
[e.g. {\it Wyss and Habermann}, 1988], the ``relative'' quiescence
[{\it Matsu'ura}, 1986; {\it Drakatos}, 2000] takes into account
  the aftershocks and detects changes in seismic activity after a
large mainshock by comparison with the usual Omori law decay of aftershocks.
{\it Drakatos} [2000] studied several aftershock sequences in Greece
which contains large aftershocks, {\it i.e.} aftershock with magnitude
no smaller than $M-1.2$, where $M$ is the mainshock magnitude.
For each sequence, he fitted the aftershock sequence by a modified
Omori law up to the time of the large aftershock using a maximum
likelihood method.
He found that large aftershocks were often preceded by a relative quiescence
by comparison with an Omori law, with an increase of the seismicity
rate just before the large mainshock occurrence.
  Such a departure from an Omori law is  predicted by our results in the
super-critical regime. Indeed, in the super-critical regime, large aftershocks
are likely to occur when the earthquake rate $N(t)$  changes from an Omori
  law to the exponential explosion for times close to $t^*$.

To illustrate this concept,
we have performed a simulation of the ETAS model in the super-critical regime
and have applied the same procedure as used by {\it Drakatos} [2000]  to fit
the synthetic aftershock sequence by an Omori law
up to the time of the first large aftershock. The parameters of the
synthetic catalog are
$K= 0.024$, $m_0= 0$, $c= 0.001$, $\alpha= 0.5$, $b= 0.8$ and $\theta=0.2$,
yielding $n=1.27$ and $t^*=4.6$.
Figure \ref{quiescence} represents the cumulative aftershock
   number as a function of time for the synthetic catalog  and the fit with
   a modified Omori law. From this figure, we see a clear relative
seismic quiescence, as defined by a cumulative aftershock number smaller
   than that predicted by the fit. The aftershock activity recovers the
   level predicted by the fit at the time of the large aftershock.
All theses results are similar to those obtained by {\it Drakatos} [2000].

In the case $n>1$, our results predict an exponential increase of the
seismicity rate at  large times.
Because we assume that the magnitude distribution is independent of time,
the same exponential acceleration is expected for both the cumulative energy
release and the cumulative number of earthquakes.
{\it Sykes and Jaum\'e} [1990] found that several large earthquakes 
in the San-Francisco
  Bay area where preceded by an acceleration of the cumulative energy 
release that
  can be fitted by an exponential function, as predicted by our results.
In laboratory experiments of rupture, several studies have also observed an
exponential acceleration of the seismic energy release before the macroscopic
rupture [{\it Scholz}, 1968; {\it Meredith et al.}, 1990; {\it Main 
et al.}, 1992].

More recently, many studies have reported an acceleration
of seismicity prior to great events (see [{\it Sammis and Sornette},
2001; {\it Vere-Jones et al.}, 2001] for reviews)
but they used a power-law instead of an exponential law
  to fit the acceleration of seismicity.
A power-law increase of the seismicity before rupture
is predicted by several statistical models of rupture in heterogeneous media,
  which consider the global rupture or the great earthquake as a critical point
  (see [{\it Sornette}, 2000a] for a review).
  Note that it is often difficult to distinguish in real data an
exponential increase from a power-law increase, especially with a small number
 of points and for times far from the rupture time. No systematic study has
  been undertaken that compares these two laws to test
if the acceleration of the seismicity is better fitted by a power-law
rather than by an exponential law (see however [{\it Johansen et al., 1996}]).

We have stressed that the ETAS model is fundamentally a mean field
approximation (branching process)
which neglects ``loops'', {\it i.e.}, multiple interactions (see Figure 1).
An important consequence
of this approximation is that the super-critical regime cannot lead to
a growth rate faster than exponential. Indeed, recall that an
exponential growth is characterized by a time derivative of the
number of events proportional to the number of events $dN/dt = N/t^*$,
 {\it i.e.}, is fundamental a {\it linear} process.
In a sequel to the present work [{\it Sornette and Helmstetter}, 2001],
we show however that
for $b<\alpha$, the impact of the largest earthquake induces an
effective nonlinearity which leads to a faster-than-exponential
growth rate, possibly leading to a finite-time singularity
[{\it Sammis and Sornette}, 2002]. A faster-than-exponential
growth rate may also be obtained
by introducing multiple interactions between earthquakes and positive feedback:
 rather than
the linear law $dN/dt = N/t^*$ expressing the condition that each
``daughter'' has only
one ``mother'', we may expect an effective law $dN/dt \sim N^{\delta}$,
with $\delta >1$
providing a measure of the effective number of ancestors impacting
directly on the
birth of a daughter.  We may thus expect that an improvement of the ETAS model beyond
the ``mean-field'' approximation would
lead to power law acceleration of seismicity.

Other precursory patterns may also be related to the super-critical
regime: they comprise the
precursory earthquake swarm or burst of aftershocks [{\it Evison}, 1977;
{\it Keilis-Borok et al.}, 1980a, 1980b; {\it Molchan et al.}, 1990;
{\it Evison and Rhoades}, 1999].
Swarms are earthquake sequences characterized by high clustering in
space and time and the occurrence of
several large events with magnitude larger than $M-1$, where
$M$ is the magnitude of the largest event. A burst of aftershocks is a sequence
   of one or more mainshocks with abnormally large number of aftershocks at the
beginning of their aftershock sequences [{\it Keilis-Borok et al.}, 1980a].
  From our results, an abnormally high aftershock rate or a sequence
with several large events are expected in the super-critical regime.

\subsection{Temporal change of $n$-value and transition from one
regime to the other one}

It is often reported that the $b$ and $p$ values vary in space and time
   [e.g., {\it  Smith}, 1981; {\it  Guo and Ogata}, 1995, 1997;
{\it Wiemer and Katsumata}, 1999].
  We  have documented that a part of the observed variation of the 
exponent $p$ may not be
genuine but result from an inadequate parameterization of a more 
complex reality.
Because $n$ and $t^*$ are function of $b$, $p$ and the other ETAS parameters,
we expect the fundamental parameters of the ETAS model, namely
$n$ and $t^*$, to vary significantly in space and time.
The branching ratio $n$ plays the role of a ``control'' parameter
quantifying the distance from the critical point $n=1$ between the sub-critical
and the super-critical regime; $t^*$ is a cross-over time and is sensitive to
details of the systems. As a consequence, it is very reasonable to expect
that the Earth's crust will change from the sub-critical to the
super-critical regime and vice-versa, as a function of time and location.

Equation (\ref{second}) shows that the branching ratio $n$ is a
decreasing function of $b$.
Accordingly, this may rationalize the observation that large earthquakes
are sometimes preceded by a decrease of the $b$-value [e.g. {\it Smith}, 1981].
A decrease of the $b$-value leads to an increase of the $n$-value,
that can move the
seismicity from the sub-critical to the super-critical regime, and
thus increase
the probability to observe a large earthquake. Other ETAS parameters
($\alpha$, $K$, $p$
and $c$) may also change in time and move the seismicity from one
regime to the other one.
{\it Ogata} [1989] measured the ETAS model parameters before and
after the 1984 Western
Nagano Prefecture earthquake ($M=6.8$).
He found that the seismic activity preceding the mainshock was characterized by
a lower $b$, $c$, $K$ parameters and local $p$ values
than the seismicity following the mainshock.
He also obtained a larger $\alpha$-value for the seismicity preceding
the mainshock.
All these changes of parameters, except the change in $K$, lead to a
larger $n$-value
before the mainshock than after. Before the mainshock, $n$ is in principle
infinite because the local $p$-value is smaller than one. As we
already discussed, this corresponds to an explosive super-critical regime
of growing seismicity. After the mainshock, we find $n=0.92$ and
$t^*=10^6$ days, using
the determination of the ETAS parameters. The seismicity has thus
changed from a super-critical regime before the mainshock to a
sub-critical regime
after the mainshock.

\section{Conclusion}

We have provided analytical solutions of the ETAS model, which describes
foreshocks, aftershocks and mainshocks on the same footing.
Each event triggers an aftershock sequence  with a rate that decays
according to the local Omori law with an exponent $p=1+\theta$.
The number of aftershocks per event increases with its magnitude.
We suggest that the Earth's crust at a given time and location may
be characterized by its branching ratio $n$, quantifying its regime.
We propose that $n$ is a fundamental parameter for understanding
and characterizing the organization of the seismicity within the Earth's crust.
In the sub-critical regime ($n<1$), the global rate of aftershocks
(including secondary aftershocks) decays with the time from the mainshock
  with a decay law different from the local Omori law.
We find  a crossover from an  Omori exponent $1-\theta$ for $t<t^*$ to
  $1+\theta$ for $t>t^*$. The modified Omori law is thus only an approximation
  of the decay law of the global aftershock sequence.
In the super-critical regime ($n>1$ and $\theta>0$), we find a novel transition from an
Omori decay law with an exponent  $1-\theta$ at early times to an
explosive exponential increase of the seismicity rate at large times.
The case $\theta<0$ leads to an infinite $n$-value, due to the slow
 decay with time of the local Omori law. In this case,
we find a transition from an Omori law with exponent
$1-|\theta|$ similar to the local law, to an exponential increase at large times, 
with a crossover time $\tau$ different from the characteristic 
time $t^*$ found in the case $\theta>0$.
These results can rationalize many of the stylized facts reported for
foreshock and aftershock sequences, such as the suggestion  that
a small $p$-value may be a precursor of a large earthquake,
the relative seismic quiescence preceding large aftershocks,
the positive correlation between $b$ and $p$-values, the observation that
great earthquakes are sometimes preceded by a decrease of $b$-value
and the acceleration of the seismicity preceding great earthquakes.


Finally, we would like to mention that our analysis can be generalized to
various other choices of the local Omori law and of the magnitude distribution.
The ETAS model can also be extended to describe the spatial
distribution of the seismicity [{\it Helmstetter and Sornette}, 2002].

\vskip 0.5cm
\acknowledgments
We are very grateful to J.R. Grasso, G. Ouillon,
V. Pisarenko, A. Sornette and D. Vere-Jones for useful suggestions and discussions and
to D. Vere-Jones as a referee for his constructive remarks and for
pointing out the relevant mathematical literature on ``self-exciting'' point processes.
We thank Y. Ogata for kindly providing a copy of Ramselaar's Master thesis.

\appendix
\section{Appendix: technical derivation of the analytical solution}

In this appendix, we provide the technical derivation of the
results used in the main text for the sub-critical and super-critical regimes.
We start from equation (\ref{third}).

\subsection{General derivation for $\theta >0$}

The integral over $\tau$ is the convolution of
$\lambda_{m'}$ with $\phi_{m'}$. Since there
is an origin of time
and we have a convolution operator, the natural tool is the Laplace transform
${\hat f}(\beta) \equiv \int_0^{+\infty} f(t) e^{-\beta t} dt$. Applying the
Laplace transform to (\ref{third}) yields
\be
{\hat \lambda}_m(\beta) =  {\hat S}(\beta,m) + P(m) \int_{m_0}^{\infty} dm' ~~
{\hat \phi}_{m'}(\beta) ~{\hat \lambda}_{m'}(\beta) ~.
\label{fourth}
\ee
where the r.h.s. has used the convolution theorem that the Laplace transform
of a convolution of two functions is the product of the Laplace
transform of the
two functions. Let us now apply the
integral operator $\int_{m_0}^{\infty} dm ~{\hat \phi}_{m}(\beta)$ on
both sides of (\ref{fourth}) and define
\be
{\cal \lambda}(\beta) \equiv \int_{m_0}^{\infty} dm ~{\hat
\phi}_{m}(\beta)~{\hat \lambda}_m(\beta)~,
\label{ghgala}
\ee
\be
Q(\beta) \equiv \int_{m_0}^{\infty} dm ~{\hat \phi}_{m}(\beta)~P(m)~,
\label{nhghgallq}
\ee
and
\be
{\cal S}(\beta) \equiv \int_{m_0}^{\infty} dm ~{\hat
\phi}_{m}(\beta)~{\hat S}(\beta,m)~.
\label{ngnhgal}
\ee
Then, expression (\ref{fourth}) yields
\be
{\cal \lambda}(\beta)  = {\cal S}(\beta) + Q(\beta) {\cal \lambda}(\beta)~,
\ee
whose solution is
\be
{\cal \lambda}(\beta)  = {{\cal S}(\beta) \over 1 - Q(\beta)}~.
\label{bghnglal}
\ee
This expression gives $\lambda_m(t)$ after inversion of the integral operator
$\int_{m_0}^{\infty} dm ~{\hat \phi}_{m}(\beta)$
and of the Laplace transform.

The key quantity controlling the dependence of $\lambda_m(t)$ is
\be
Q(\beta) = {K \over \theta c^{\theta}}
\left(\int_{m_0}^{\infty} dm ~10^{\alpha(m-m_0)} ~P(m)\right)
~\left( \theta \int_0^\infty dt {e^{-\beta c t} \over
(t+1)^{1+\theta}}\right)~,
\label{jgmjslqlqllq}
\ee
obtained by replacing the expression of $\phi_m(t)$ defined in (\ref{first})
and normalizing $t/c \to t$.
Using $P(m) = \ln(10) ~ b ~ 10^{-b(m-m_0)}$, we obtain
\be
Q(\beta) = n ~R(\beta c),
\ee
where we have used the expression (\ref{second}) of $n$ and defined
\be
R(\beta) \equiv \theta \int_0^\infty dt {e^{-\beta t} \over (t+1)^{1+\theta}}
= \theta~ e^{\beta}~\beta^{\theta}~\Gamma(-\theta , \beta)
= 1 - e^{\beta}~\beta^{\theta}~\Gamma(1-\theta , \beta)~,
\label{hghigwnv}
\ee
where
\be
\Gamma (a,x)= \int_x^{\infty} dt~e^{-t}~t^{a-1}\label{bhgnfala}
\ee
is the (complementary) incomplete Gamma function [{\it Abramowitz and
Stegun}, 1964] and we have used $\Gamma(1+a, x)= a \Gamma(a, x) + x^a ~e^{-x}$ obtained
by integration by part.
Using the expansion of the incomplete Gamma function  [{\it Olver}, 1974]
\be
\Gamma(a, x) = \Gamma(a) - \sum_{k=0}^{+\infty}
{(-1)^k ~ x^{a+k} \over k! ~ (a+k)} ~,   ~~~~{\rm for}~~a>0~,
\label{ngnlqlka}
\ee
we obtain
\be
R(\beta) = 1 - \Gamma(1-\theta)~\beta^{\theta} + {1 \over 1-\theta}~\beta
+ {\cal O}(\beta^{1+\theta}, \beta^{2},\beta^{2+\theta}, \beta^{3},...)~.
   \label{gnngal}
\ee
It is possible, using the full expansion of the incomplete Gamma function,
 to estimate the value of $\lambda(\beta)$ when the
second term ${1 \over 1-\theta}~\beta$ of the expansion cannot be
neglected anymore
compared with the term proportional to $\beta^\theta$.
Thus, the expansion (\ref{gnngal}) using the first two terms only
$R(\beta) = 1 - \Gamma(1-\theta) ~\beta^{\theta}$ becomes invalid for $\beta >
[\Gamma(1-\theta) (1-\theta)]^{1/(1-\theta)}$,
{\it i.e.}, for times smaller than  $[\Gamma(2-\theta)]^{-1/(1-\theta)}$. For all practical
purpose, this is a small value and we can use safely the expansion
(\ref{gnngal})
in the following calculations.

Let us now make explicit ${\cal \lambda}(\beta)$:
\be
{\cal \lambda}(\beta) = {K \over \theta c^\theta}~R(\beta c)~ \int_{m_0}^\infty dm~
10^{\alpha (m-m_0)} \int_0^\infty dt~\lambda_m(t)~e^{-\beta t}~.   \label{nngal}
\ee

Using the definition of $\lambda(t)$ given by (\ref{ggnnlalaq}) and
the factorization of the times and magnitudes in (\ref{nngal}), we obtain
\be
{\cal \lambda}(\beta) = n R(\beta c) {\hat \lambda}(\beta)~,  \label{nhglgnlaa}
\ee
where
\be
{\hat \lambda}(\beta) =  \int_0^\infty dt~\lambda(t)~e^{-\beta t}~.   \label{nngaaal}
\ee
Replacing (\ref{nhglgnlaa}) in (\ref{bghnglal}) gives
\be
{\hat \lambda}(\beta) = {{\cal S}(\beta) \over n R(\beta c) \left(1 - n
R(\beta c)\right)}~.
\label{bghnglaaal}
\ee

When a great earthquake
occurs at the origin of time $t=0$ with magnitude $M$,
$S(t,m)=\delta(t)~\delta(m-M)$, expression (\ref{ngnhgal}) gives
\be
{\cal S}(\beta) = {K \over \theta c^{\theta}}~10^{\alpha (M-m_0)}~R(\beta c)~.
\label{ngnkkee}
\ee
Thus, expression (\ref{bghnglaaal}) becomes
\be
{\hat \lambda}(\beta) = {b-\alpha \over b}~{10^{\alpha (M-m_0)} \over
\left(1 - n R(\beta c)\right)}~.
\label{bghqqaaal}
\ee


The dependence of ${\hat \lambda}(\beta)$ on $\beta$ is
uniquely controlled by the denominator $1 - n R(\beta c)$.

\subsection{The sub-critical regime $n<1$}

The analysis proceeds exactly as in [Sornette and Sornette, 1999].
For $0 < \theta < 1$, and for small $\beta$ (large times),
${\hat \lambda}(\beta)$ given by (\ref{bghqqaaal}) is
\be
{\hat \lambda}(\beta) = {S_0 \over 1 - n [1 - d (\beta c)^{\theta}]}
= {{ S_0 \over (1-n)} \left( 1 \over 1 + (\beta t^*)^{\theta}\right)},
\label{eal}
\ee
where $t^*$ is defined by (\ref{ngnvl}) and the external source term $S_0$ is
defined by  (\ref{S02}).
We retrieve equation (13) of [{\it Sornette and Sornette}, 1999] with
the correspondence $t_0 \to c$.

Two cases must be distinguished.

$\bullet$ $ \beta  t^* < 1$ corresponds to $t > t^*$
by identifying as usual the dual variable $\beta$ to $t$ in the Laplace
transform with $1/t$. In this case, we can expand
$ 1 \over 1 + (\beta t^*)^{\theta} $, which leads to
\be
{\hat \lambda}_{t>t^*}(\beta) \sim {S_0\over 1-n} [1 -  (\beta t^*)^{\theta}].
\label{fgh}
\ee
We recognize the Laplace transform of a power law of exponent
$\theta$, {\it i.e.}
\be
\lambda_{t>t^*}(t) \sim { S_0  \over \Gamma(\theta) (1-n)}~~
{ {t^*}^{\theta}\over t^{1+\theta}} ~~~~~ \mbox{  for $t>t^*$  ~.}
\label{eaz}
\ee

$\bullet$ For $t < t^*$, $\beta t^* > 1$ and  (\ref{eal})
can be written with a good approximation as
\be
{\hat \lambda}_{t<t^*}(\beta) = {S_0 \over  (1-n) (\beta t^*)^{\theta}}
   \sim \beta ^{-\theta} .
\label{elkkk}
\ee
Denoting $\Gamma(z) \equiv \int_0^{+\infty} dt~e^{-t} ~t^{z-1}$, we see that
$\int_0^{+\infty} dt~e^{-\beta t}~ t^{z-1} = \Gamma (z) \beta^{-z}$. Comparing
with (\ref{elkkk}), we thus get
\be
\lambda_{t<t^*}(t) \sim { S_0 \over \Gamma(\theta )(1-n) }~~~
{{t^*}^{-\theta} \over t^{1-\theta}} ~~~~~ \mbox{  for $t<t^*$  .}
\label{eadfgz}
\ee

We verify the self-consistency of the two solutions $\lambda_{t>t^*}(t)$
and $\lambda_{t<t^*}(t)$
by checking that $\lambda_{t>t^*}(t^*) = \lambda_{t<t^*}(t^*)$. In other words,
$t^*$ is indeed the transition time at which the ``short-time''
regime $\lambda_{t<t^*}(t)$ crosses over to the ``long-time'' regime $\lambda_{t>t^*}(t)$.

We now calculate the full expression of $\lambda(t)$ valid at all times.
We expand
\be
{1 \over (\beta  t^*)^{\theta}+1}
   = {1 \over (\beta t^*)^{\theta}}~
{1 \over (\beta t^*)^{-\theta} +1}
= {1 \over (\beta t^*) ^{\theta}}
~\sum_{k=0}^{\infty}(-1)^k (\beta t^*) ^{-k \theta}~,
\ee
Thus, by taking the inverse Laplace transform
\be
\lambda(t) = {S_0 \over 1-n }~{1 \over 2\pi i}~
\int_{c-i\infty}^{c+i\infty}d\beta~e^{\beta t}
~ \sum_{k=0}^\infty (-1)^k (\beta t^*) ^{- (k+1) \theta} ~.
\ee
The inverse Laplace transform of $\beta^{-(k+1)\theta}$ is
$t^{(k+1)\theta -1}/\Gamma((k+1)\theta)$.
This allows us to write
\be
\lambda(t) = {S_0 \over 1-n}~{{t^*}^{-\theta} \over t^{1-\theta}}~
\sum_{k=0}^\infty (-1)^k {(t/t^*)^{k\theta} \over
\Gamma((k+1)\theta)}
\label{bhgknaalkal1}
\ee
Expression (\ref{bhgknaalkal1}) provides the solution that describes
the cross-over from the $1/t^{1-\theta}$ Omori's law (\ref{eadfgz})
at early times to
the $1/t^{1+\theta}$ Omori's law (\ref{eaz}) at large times.
The series $\sum_{k=0}^\infty (-1)^k {(t/t^*)^{k\theta} \over
\Gamma((k+1)\theta)}$ is a series representation of the special Fox function
[{\it Gl\"ockle and Nonnenmacher}, 1993]
and it is also related to the generalized Mittag-Leffler function.

For large times $t>>t^*$, a direct numerical evaluation of $\lambda(t)$ from equation
 (\ref{bhgknaalkal1}) is impossible due to the very slow convergence
  of the series. The pad\'e summation method
  [{\it Bender and Orzag}, 1978] can be used to improve the convergence of this series
  and to evaluate numerically (\ref{bhgknaalkal1}) for all times.


\subsection{The super-critical regime $n>1$}

We can analyze this regime by putting $n>1$ in (\ref{bghqqaaal}) which can be
written under a form similar to (\ref{eal}):
\be
{\hat \lambda}(\beta) = {S_0 \over \left(1 - n R(\beta c)\right)}
= {S_0 \over d n (\beta c)^{\theta} - (n-1)}
= {{ S_0 \over (n-1)} \left( 1 \over (\beta t^*)^{\theta}-1 \right)},
\label{bssghaqqaaal}
\ee
In the second and third equalities of (\ref{bssghaqqaaal}),
we have used the small $\beta$-expansion
(\ref{gnngal}) of $R(\beta c)$ valid for $0 < \theta < 1$.

At  early times $c \ll t \ll t^*$, {\it i.e.}, $\beta t^* \gg 1$,
${\hat \lambda}(\beta) \approx {S_0 \over (n-1) (\beta t^*)^{\theta}}$
which is the Laplace transform of (\ref{eadfgz}): thus, the 
early decay rate of
aftershocks is the same $\sim 1/t^{1 - \theta}$ as for the
sub-critical regime (\ref{eadfgz}).
However, as time increases, the dual $\beta$ of $t$ decreases
and ${\hat \lambda}(\beta)$ grows faster
than $\sim (\beta c)^{-\theta}$ due to the presence of the negative
term $- (n-1)$.
This can be seen as an apparent exponent $\theta_{\rm app}>\theta$
increasing progressively such that
$d n (\beta t^*)^{\theta} - 1 \approx C (\beta t^*)^{\theta_{\rm
app}}$, where $C$ is a constant.
Note that $\theta_{\rm app}>\theta$ for the pure power law $C (\beta
c)^{\theta_{\rm app}}$
to mimic the acceleration induced by the negative correction
$- (n-1)$. The seismic rate will thus decay approximately as $\sim
1/t^{1 - \theta_{\rm app}(t)}$.

The large time behavior is controlled by the pole at $\beta=1/t^*$ of
${\hat \lambda}(\beta)$. Close to $1/t^*$,
\be
{\hat \lambda}(\beta) \approx {S_0 \over (n-1) \theta  }
~  {1 \over \beta t^* - 1 }~.
\label{bssgweel}
\ee
The inverse Laplace transform is thus
\be
\lambda(t) =(2\pi i)^{-1} \int_{c-i\infty}^{c+i\infty}
d\beta~e^{\beta t} ~{\hat \lambda}(\beta)
\sim {S_0 \over (n-1) t^* \theta} e^{t/t^*}  \label{gnalksd}
\ee
exhibiting the exponential growth at large times. Expression (\ref{ngnvl})
shows that $1/t^* \sim |1-n|^{1 \over \theta}$. Thus, as expected,
the exponential growth
disappears as $n \to 1^+$.

We now calculate the full expression of $\lambda(t)$ valid at all times.
We expand
\be
{1 \over (\beta  t^*)^{\theta}-1}
   = {1 \over (\beta t^*)^{\theta}}~
{1 \over 1-(\beta t^*)^{-\theta} }
= {1 \over (\beta t^*) ^{\theta}}
~\sum_{k=0}^{\infty} (\beta t^*) ^{-k \theta}~,
\ee

Thus
\be
\lambda(t) = {S_0 \over (n-1) }~{1 \over 2\pi i}~
\int_{c-i\infty}^{c+i\infty}
d\beta~e^{\beta t} ~ \sum_{k=0}^\infty {(\beta t^*)^{- (k+1) \theta}} ~.
\ee
The inverse Laplace transform of $1/\beta^{(k+1)\theta}$ is
$t^{(k+1)\theta -1}/\Gamma((k+1)\theta)$.
This allows us to write
\be
\lambda(t) = {S_0 \over (n-1) }~{{t^*}^{-\theta} \over t^{1-\theta}}~
\sum_{k=0}^\infty {(t/t^*)^{k\theta} \over \Gamma((k+1)\theta)}
\label{bhgknaalkal}
\ee
Expression (\ref{bhgknaalkal}) provides the solution that describes
the cross-over from the $1/t^{1-\theta}$ Omori's law at early times to
the exponential growth at large times. Note that the solution
(\ref{bhgknaalkal})
can be obtained directly from (\ref{bhgknaalkal1}) by removing the
alternating sign
$(-1)^k$ in the sum. The solution (\ref{bhgknaalkal}) retrieves the two
regimes discussed before.
\begin{enumerate}
\item For $t<t^*$, the sum in (\ref{bhgknaalkal}) is close to
$1/\Gamma(\theta)$, which leads to
\be
\lambda(t) \approx {S_0 \over \Gamma(\theta) (n-1)}~{{t^*}^{-\theta} \over
t^{1-\theta}}~.  \label{gnlanvlaa}
\ee
\item For $t \geq t^*$, the sum dominates. The sum is very similar to
the series expansion of $e^{t/t^*}$ and is actually proportional to
$e^{t/t^*}$ for large $t$. This result is obvious for $\theta=1$
since the series expansion becomes identical to that of
$e^{t/t^*}$. This can be justifies for other values of $\theta$ as follows.
For $\theta \to 0$,
the discrete sum transforms into a continuous integral of the type
\be
\int_0^{\infty} dx ~t^x/\Gamma(x)~.
\label{gndvnqlq}
\ee
A saddle-node approximation, performed using the
Stirling approximation (which already gives a very good precision for
small $z$) $\Gamma(z)  \approx \sqrt{2\pi}~ e^{-z}~z^{z-{1 \over
2}}$, shows that the saddle-node of the
integrant occurs for $x\approx t/t^*$, which then gives
$\lambda(t) \sim e^{t/t^*}$. For arbitrary $\theta$, we can use the
Poisson's summation rule
\be
\sum_{r=-\infty}^{+\infty} f(r) = \int_{-\infty}^{+\infty}  du~f(u)~+~
\sum_{q=1}^{+\infty} \int_{-\infty}^{+\infty}  du~f(u)~\cos [2\pi q u]~,
\label{ngnlalqa}
\ee
on the function defined by
\be
f(r) \equiv {(t/t^*)^{r\theta } \over
\Gamma(r\theta+\theta)}~,~~~~{\rm for~} r \geq 0
\ee
and $f(r)=0$ for $r < 0$. The left-hand-side of (\ref{ngnlalqa}) is nothing
but the semi-infinite sum in (\ref{bhgknaalkal}). The first term in
the right-hand-side
retrieves the integral (\ref{gndvnqlq}) encountered for the case
$\theta \to 0$.
This term thus contributes a term proportional to $e^{t/t^*}$.
All the other terms contribute negative powers of $t$ and are thus
negligible compared
to the exponential for $t > t^*$. This can be seen from the fact that each term
with $q \geq 1$ is similar to the sum in (\ref{bhgknaalkal1}) for the
subcritical
case with alternating signs. The larger $q$ is, the faster is the frequency of
The leading dependence $\lambda(t) \sim e^{t/t^*}$ valid for any $0 \leq
\theta \leq 1$
retrieves the limiting behavior
already given in (\ref{gnalksd}) from a different approach for large
times $t>>t^*$. It has also been proved rigorously in 
[{\it Ramselaar}, 1990].

\end{enumerate}

\subsection{Case $\theta <0$ corresponding to a local Omori's law
exponent $p<1$}

The general equation (\ref{third}) still holds in this case and the general
derivation starting with (\ref{fourth}) up to (\ref{bghnglal}) still applies.
The key quantity controlling the dependence of $\lambda_m(t)$ is still
$Q(\beta)$ defined by (\ref{jgmjslqlqllq}). Writing $\theta =
-|\theta|$, we have
\be
Q(\beta) = n_0 ~R'(\beta c),
\ee
where $n_0$ is defined by (\ref{demmfmlwwq}) and
\be
R'(\beta) \equiv \int_0^\infty dt {e^{-\beta t} \over (t+1)^{1-|\theta|}}
= e^{\beta}~\beta^{-|\theta|}~\Gamma(|\theta| , \beta)
\label{hghigwnaav}
\ee
where $\Gamma (a,x)$ is the (complementary) incomplete Gamma function defined
by (\ref{bhgnfala}). Using the exact expansion (\ref{ngnlqlka}), we obtain
\be
Q(\beta) = n_0 ~e^{\beta c}~(\beta c)^{-|\theta|}~
\left( \Gamma(|\theta|) - \sum_{k=0}^{+\infty}
{(-1)^k ~ (\beta c)^{|\theta|+k} \over k! ~ (|\theta|+k)} \right) ~.
\label{gnhgnw}
\ee
For small $\beta$'s ({\it i.e.}, large times), expression (\ref{gnhgnw}) has the
following leading behavior
\be
Q(\beta) = n_0~\Gamma(|\theta|) ~(\beta c)^{-|\theta|}~-~ {n_0 \over |\theta|}
~+~ n_0 \Gamma(|\theta|) ~(\beta c)^{1-|\theta|} + h.o.t.
\label{bhhntjgjg}
\ee
where $h.o.t.$ stands for higher-order terms in the expansion in increasing
powers of $\beta c$.

The source term ${\cal S}(\beta)$ in the denominator of ${\hat \lambda}(\beta)$
given by (\ref{bghnglal}) is now given by
\be
{\cal S}(\beta) = K~ c^{|\theta|}~10^{\alpha (M-m_0)}~R'(\beta c)~.
\label{ngnkkeaae}
\ee
Expression (\ref{bghnglal}) for ${\hat \lambda}(\beta)$ then yields
\be
{\hat \lambda}(\beta) = {S_0 \over 1 - Q(\beta c)}~,
\label{bghqqadvfvdaal}
\ee
where $R'(\beta c)$ is given by (\ref{hghigwnaav}), $n_0$ is
defined by (\ref{demmfmlwwq}) and $S_0$ is defined by (\ref{S02}).
The dependence of ${\hat \lambda}(\beta)$ on $\beta$ is
uniquely controlled by the denominator $1-Q(\beta c) = 1 - n_0 R'(\beta c)$.

Using (\ref{bhhntjgjg}), we get the leading behavior for small $\beta c$
\ba
{\hat \lambda}(\beta) &=& {S_0 \over
1 + {n_0 \over|\theta|} - n_0 \Gamma(|\theta|) ~(\beta
c)^{-|\theta|}}\nonumber \\
&=& {S_0 \over (1+{n_0 \over|\theta|})}
{1 \over (1 - (\beta \tau)^{- |\theta|})}
\label{bghqqadsqvfvdaal}
\ea
where the characteristic time $\tau$ is given by (\ref{jgjjgswsa}).

At early times $c<t<\tau$, $(\beta \tau)^{-|\theta|}<1$ so that
\be
{\hat \lambda}(\beta) \approx {S_0 \over (1 + {n_0 \over|\theta|})} ~
\left(1 + (\tau \beta)^{-|\theta|} \right)~.
\label{jgjjgswsa2}
\ee
By applying the inverse Laplace transform, the constant term contributes
a Dirac function $\delta(t)$ which is irrelevant as the calculation is valid
only for $t>c$. The other term $(\tau \beta)^{-|\theta|}$ gives
\be
\lambda(t)= {S_0 \over (1 + {n_0 \over|\theta|}) \Gamma(|\theta|)} ~
{ \tau ^{-|\theta|} \over t^{1-|\theta|}}~.
\label{jgjjgswsa3}
\ee
The early time behavior of $\lambda(t)$ is thus similar to the local Omori law $1 / t^{1-|\theta|}$.

Similarly to the super-critical case $n>1$ of the regime $\theta >0$, the
long time dependence of the regime $\theta <0$ is controlled by a simple pole
$\beta^*={1 \over \tau}$.

Thus, the long-time seismicity is given by
\be
\lambda(t) = {S_0 \over (1 + { n_0 \over |\theta|}) \tau |\theta|}  ~ e^{t/\tau}
\label{gnalksd2}
\ee

We can also calculate the full expression of $\lambda(t)$ valid at all times $t>c$.
We expand
\be
{1 \over 1- (\beta  \tau)^{-|\theta|}}
= \sum_{k=0}^{\infty} (\beta \tau) ^{-k |\theta|}~,
\ee
Removing the constant term, which by the inverse Laplace transform contributes
a Dirac function $\delta(t)$ which is irrelevant as the calculation is valid
only for $t \gg c$, we get
\be
\lambda(t) = {S_0 \over (1 + { n_0 \over |\theta|}) }~{1 \over 2\pi i}~
\int_{c-i\infty}^{c+i\infty}
d\beta~e^{\beta t} ~ \sum_{k=1}^\infty {(\beta \tau)^{- k |\theta|}} ~.
\ee
The inverse Laplace transform of $1/\beta^{k|\theta|}$ is
$t^{k|\theta| -1}/\Gamma(k|\theta|)$.
This allows us to write
\be
\lambda(t) = {S_0 \over (1 + { n_0 \over |\theta|}) }~{1 \over t}~
\sum_{k=1}^\infty {(t/\tau)^{k|\theta|} \over \Gamma(k|\theta|)}
\label{bhgknaalkal3}
\ee
Expression (\ref{bhgknaalkal3}) provides the solution that describes
the cross-over from the local Omori law $1/t^{1-|\theta|}$ at early times to
the exponential growth at large times.

\end{article}

\clearpage

\clearpage
\begin{planotable}{llccccccccc}
\tablewidth{44pc}
\tablenum{1}
\label{tab1}
\tablecaption{ ETAS parameters, branching ratio $n$ and
characteristic time $t^*$  for
the sequences studied by {\it Ogata} [1989, 1992].
We have computed $n$ and $t^*$ using equations (\ref{second}) and (\ref{ngnvl})
from the ETAS parameters $K$, $\alpha$, $c$, $p=1+\theta$ and $\mu$
calculated by {\it Ogata} [1989, 1992] using a maximum likelihood method.
For most sequences, we have assumed $b=1$ to evaluate $n$ and $t^*$ because
$b$-value is not given in [{\it Ogata}, 1989, 1992].
Thus, there is a large uncertainty in the $n$ and $t^*$ values
 in the case where $\alpha$ is close to 1.}
\tablehead{\colhead{Reference} & \colhead{seismicity data} & \colhead{$M_0$}
& \colhead{$b$} & \colhead{$\mu$} & \colhead{$K$}
  &\colhead{ $c$}  & \colhead{$p$}  & \colhead{$\alpha$}
   & \colhead{$n$}  & \colhead{$t^*$}  \\
   & & & & \colhead{day$^{-1}$} & & \colhead{day} & & & & \colhead{day}   }
\tablenotetext{a}{$t^*$ cannot be evaluated because $p=1$}
\tablenotetext{b}{$t^*$ cannot be evaluated because $\alpha<b$}
\tablenotetext{c}{$\tau$ is given instead of $t^*$ because $\theta<0$}

\startdata

{\it Ogata} [1989] & Japan, 1895-1980
& 6.0 & 1.0  & 0.005 &  0.087 &  0.02  &  1.0  & 0.7  & Inf & \tablenotemark{a} \nl

{\it Ogata} [1989] & Rat-Island 1963-1982
& 4.7    & 1.0  & 0.0  &  0.072 &  0.167 &  1.35 & 0.63 & 1.04 & $4600$\nl

{\it Ogata} [1989] & Nagano, 1978-1986
& 2.5 & 0.9  & 0.021 &  0.008& 0.017  &  0.85 & 0.94 & Inf & \tablenotemark{b}  \nl

{\it Ogata} [1989] & Nagano aftershocks, 1986
& 2.9 & 1.2  & 0.0   &  0.032 &  0.038 & 1.14  & 0.73 & 0.92 &$4.10^6$\nl

{\it Ogata} [1992] & worldwide shallow earthquakes
& 7.0  & 1.0   &   0.019   &  0.018  &   0.21  &  1.03  &  0.53  & 1.49 &  $10^{17}$ \nl

{\it Ogata} [1992] & Central Aleutian, 10 years
& 4.7 & 1.0  & 0.008 &  0.042 & 0.03  & 1.13 & 0.62 & 1.34 & $2200$\nl

{\it Ogata} [1992] & Tohoku, 95 years
&6.0  & 1.0 &   0.0054  & 0.98  &  0.02 & 1.0    & 0.70  & Inf & \tablenotemark{a}\nl

{\it Ogata} [1992] & Tokachi-Oki aftershocks, 1 year
&4.8&   1.0 &   0.14  & 0.015  & 0.23  & 1.28 &  0.98 &4.03	&1.5\nl

{\it Ogata} [1992] & Niigata aftershocks, 150 days
&4.0  & 1.0 &   0.075  & 0.0005 &  0.15 & 1.37  &1.26 & Inf & \tablenotemark{b}\nl

{\it Ogata} [1992] & Niigata aftershocks, 150 days
&2.5   &1.0   & 0.47  & 0.0002 &  1.10  & 1.72  & 1.34 & Inf & \tablenotemark{b}\nl

{\it Ogata} [1992] &Izu Islands, 55 years
&4.0  & 1.0  &  0.0038 &   0.062  &  0.012 &  1.143 &  0.155 & 0.96  &  $10^{8}$ \nl

{\it Ogata} [1992] &Izu Peninsula, 7 years
& 2.5 &   1.0  &  0.022  & 0.035 &  0.003 & 1.35 & 0.17 & 0.91 &   7.3 \nl

{\it Ogata} [1992] & Off east cost of Izu, 33 days
&2.9 &  1.0  &  0.59 &  0.016  & 0.009 & 1.73 & 0.31 &1.00&  346.\nl

{\it Ogata} [1992] & Matsushiro swarm, 20 years
&3.9 &  1.0  &  0.0006  & 0.092  & 0.13 &  1.14  & 0.27  & 1.21  &  2200 \nl

{\it Ogata} [1992] & Kanto, 1904-1916
& 5.4 & 1.0  & 0.028    & 0.010  &  0.010 &  1.00 & 0.62 & Inf & \tablenotemark{a} \nl

{\it Ogata} [1992] & Kanto, 1916-1923
& 5.4 & 1.0  & 0.025    & 0.001  &  0.010 &  1.02 & 1.31 & Inf & \tablenotemark{b} \nl

{\it Ogata} [1992] & Hachijo, 1938-1969
 & 5.4 & 1.0  & 0.013 & 0.008  &  0.004 & 1.02 & 0.85 & 3.0 &  $5.10^6$\nl 
{\it Ogata} [1992] & Hachijo, 1969-1973
 & 5.4 & 1.0  & 0.016 & 0.001  &  0.013 & 1.00 & 1.11 & Inf & \tablenotemark{a}\nl 

{\it Ogata} [1992] & Tonankai, 1933-1939
& 5.2 & 1.0  & 0.050 &  0.010 &  0.065 & 1.02 & 0.90 & 5.28 & $4.10^3$\nl
{\it Ogata} [1992]  & Tonankai, 1939-1944
& 5.2 & 1.0  & 0.031 &  0.009 & 0.011  & 1.01 & 0.83 & 5.54 & $10^7$ \nl

{\it Ogata} [1992]  & Tokachi, 1926-1945
 & 5.0 & 1.0  & 0.047 & 0.013  &  0.065 & 1.32 & 0.83 & 0.57 & 0.40 \nl
{\it Ogata} [1992]  & Tokachi, 1945-1952
 & 5.0 & 1.0  & 0.041 & 5.20  & 11.6  & 3.50 & 1.37 & Inf &  \tablenotemark{b}   \nl 
{\it Ogata} [1992] & Tokachi, 1952-1961
 & 5.0 & 1.0  & 0.032 &  0.021 &  0.059 & 1.10 & 0.72 & 0.99 & $10^{22}$\nl
{\it Ogata} [1992]  & Tokachi, 1961-1968
 & 5.0 & 1.0  & 0.014 & 0.014  & 0.005  & 0.86 & 0.43 & Inf & $7.10^{5}$\tablenotemark{c}
\end{planotable}
\end{document}